\documentclass[usenatbib]{mn2e}

\usepackage{empheq}

\usepackage{graphicx}
\usepackage{txfonts}
\usepackage{bm}
\usepackage{natbib}

\usepackage{hyperref}
\hypersetup{colorlinks=true,
            citecolor=blue,
            linkcolor=blue}

\usepackage{listings}           

\def\vv#1{\bm{#1}}
\def\be {\begin{equation}}
\def\ee {\end{equation}}
\def\vR {\vv{R}}
\def\ddvR{\vv{\ddot R}}
\def\vf {\vv{f}}
\def\vg {\vv{g}}
\def\vh {\vv{h}}
\def\vu {\vv{u}}

\def\vr {\vv{r}}
\def\vrkl {\vv{r}_{kl}}
\def\ur {\vv{\hat r}}
\def\urkl {\vv{\hat r}_{kl}}
\def\rkl {r_{kl}}

\def\vF{\vv{F}}
\def\vT {\vv{T}}

\def\ddvr{\vv{\ddot r}}
\def\ddvR{\vv{\ddot R}}
\def\omt{\omega}
\def\w{\Omega}
\def\vw {\vv{\w}}
\def\tvw {\vv{\tilde \w}}
\def\vL {\vv{L}}
\def\vK {\vv{K}}
\def\dvL{\vv{\dot L}}

\def\ia{1}
\def\jb{2}
\def\kc{3}
\def\vi {\vv{i}}
\def\vj {\vv{j}}
\def\vk {\vv{k}}
\def\xii {\hat x}
\def\xj {\hat y}
\def\xk {\hat z}

\def\TA{{\cal A}}
\def\TI {{\bf \cal I}}
\def\TIv {{\bf \cal I}_\nu}
\def\TJ {{\bf \cal J}}
\def\TP {{\bf \cal K}}
\def\DR{\Delta \R}

\def\Iv{I}
\def\SR {{\bf \cal S}}

\def\M {m_0}
\def\m {m}

\def\tt {\{t\}}
\def\ep {\mathrm{e}}

\def\eq {{e}}
\def\xix{\xi}

\def\hf{h_\mathrm{f}}

\def\kf{k_\mathrm{f}}

\def\taub{\tau}

\def\nue{\eta}

\def\crm{\cr\noalign{\medskip}}

\def\X{X}
\def\Y{Y}
\def\Z{Z}
\def\R{R}

\def \Frac#1#2{{{\displaystyle\strut#1}\over{\displaystyle\strut#2}}}

\def\llabel#1{\label{#1}}
\def\nul#1{}

\def\figpath{}

\newcommand*\widefbox[1]{\fbox{\hspace{2em}#1\hspace{2em}}}

\begin{document}
%
\title[Tidal dynamics of multi-body extrasolar systems: \texttt{TIDYMESS}]{A direct $N$-body integrator for modelling the chaotic, tidal dynamics of multi-body extrasolar systems: \texttt{TIDYMESS}}


\author[T. C. N. Boekholt and A. C. M. Correia]{Tjarda C. N. Boekholt$^{1}$\thanks{E-mail: tjarda.boekholt@physics.ox.ac.uk (TB); acor@uc.pt (AC)} and Alexandre C. M. Correia$^{2,3}$\\
$^{1}$Rudolf Peierls Centre for Theoretical Physics, Clarendon Laboratory, Parks Road, Oxford, OX1 3PU, UK \\
$^{2}$CFisUC, Departamento de F\'isica, Universidade de Coimbra, 3004-516 Coimbra, Portugal \\
$^{3}$IMCCE, Observatoire de Paris, PSL Universit\'e, 77 Av. Denfert-Rochereau, 75014 Paris, France
}

\date{\today}

\maketitle

\begin{abstract}
Tidal dissipation plays an important role in the dynamical evolution of moons, planets, stars and compact remnants. The interesting complexity originates from the interplay between the internal structure and external tidal forcing. Recent and upcoming observing missions of exoplanets and stars in the Galaxy help to provide constraints on the physics of tidal dissipation. It is timely to develop new $N$-body codes, which allow for experimentation with various tidal models and numerical implementations. 
We present the open-source $N$-body code \texttt{TIDYMESS}, which stands for ``TIdal DYnamics of Multi-body ExtraSolar Systems''. This code implements a creep deformation law for the bodies, parametrized by their fluid Love numbers and fluid relaxation times. Due to tidal and centrifugal deformations, we approximate the general shape of a body to be an ellipsoid. We calculate the associated gravitational field to quadruple order, from which we derive the gravitational accelerations and torques. The equations of motion for the orbits, spins and deformations are integrated directly using a fourth-order integration method based on a symplectic composition. We implement a novel integration method for the deformations, which allows for a time step solely dependent on the orbits, and not on the spin periods or fluid relaxation times. This feature greatly speeds up the calculations, while also improving the consistency when comparing different tidal regimes. We demonstrate the capabilities and performance of \texttt{TIDYMESS}, particularly in the niche regime of parameter space where orbits are chaotic and tides become non-linear. 
\end{abstract}

\begin{keywords}
methods: numerical -- stars: kinematics and dynamics -- stars: rotation -- planets and satellites: dynamical evolution and stability. 
\end{keywords}


\section{Introduction}\label{sec:intro}

\subsection{Pure $N$-body problem}

The $N$-body problem consists of some initial configuration of $N$ bodies, represented as point-particles, and calculating their orbits in time by solving Newton's equations of motion and law of gravitation \citep{Newton_1687}. This challenging task remains an active field of research, and has resulted in numerous different $N$-body codes in the last few decades. This repertoire of codes consists of variations in the balance between accuracy and speed, long term conservation properties, and regimes of applicability. The pure $N$-body problem as described above, provides a first-order description of the dynamical evolution of planetary and stellar systems. Interesting dynamical phenomena in astronomical systems have already been analysed this way, such as exponential sensitivity and chaos \citep[e.g.][]{Poincare_1905, Miller_1964, Sussman_Wisdom_1992, Goodman_1993, TB16, Stone_2019, Boekholt_2020, spz22}, orbital resonances in planetary systems \citep[e.g.][]{Laplace_1784, Peale_1976a, Wisdom_1982, Tsiganis_etal_2005, Laskar_Gastineau_2009, Rein_2012}, Lidov-Kozai cycles in triple systems \citep[e.g.][]{Lidov_1961, Kozai_1962, Ford_etal_2000, Perets_2012, Naoz_2016, Toonen_2020}, core collapse in star clusters \citep[e.g.][]{McMillan_1990, Makino_1996, Baumgardt_2002, Tanikawa_2012, Fujii_2014}, and much more. 

Detailed observations of the moons and planets in the Solar System however, as well as recent state of the art observations of exoplanets and gravitational waves, have presented some limitations of the pure $N$-body problem. The precession of Mercury's orbit by 43 arcseconds per century has a relativistic origin \citep{Einstein_1915}, and can be reproduced by adding Post-Newtonian terms to the accelerations \citep[e.g.][]{Will_2014}. The observation that the Moon is receding from Earth by about 4 centimeters per year \citep{Dickey_etal_1994} motivated the study and addition of tides to the $N$-body problem. The population of hot Jupiters with orbital periods of a few days is hard to explain using conservative $N$-body simulations, and instead, is thought to result from planet-planet scattering or Lidov-Kozai cycles combined with tidal dissipation \citep[e.g.][]{Wu_Murray_2003, Fabrycky_Tremaine_2007, Correia_etal_2011, Beauge_Nesvorny_2012, Bataille_etal_2018}. Another example is the observed rate of gravitational wave sources, which is higher than expected from pure $N$-body simulations of stellar systems \citep[e.g.][]{Samsing_2018, Kocsis_2022}. Also here, the inclusion of gravitational tides, in combination with gravitational wave dissipation, is thought to be a fruitful line of inquiry \citep[e.g.][]{Bildsten_1992, Mandel_2016}.

\subsection{$N$-body problem with Creep tides}

For increasingly dense multi-body systems the point-particle approximation will eventually start to break down. Structural properties of the bodies themselves have to be taken into account. These typically include the radius of the body, the internal density distribution (uniform or centrally concentrated), the spin rate and orientation, and the tidal response, i.e. how are the bodies tidally deformed, and how does that affect the orbits and spins? The physics of tidal dissipation is complex due to the interplay between internal structure processes, and external tidal forcing due to other bodies. The $N$-body environment is dynamic and often chaotic, thus making it challenging to calculate tidal effects and to predict the long-term tidal evolution. 

There are numerous tidal models in the literature, motivated by the wide variety of applications related to tides. The tidal response depends on the type of body, e.g. small rocky bodies, giant planets, low-mass stars, giant stars, compact remnants etc. However, models are also distinguished by whether one is interested in the effects of tides on the detailed internal structure, or rather on the orbital evolution, which affects the structure of the $N$-body system as a whole. 
Numerically, the most accurate method is to resolve each body by its own ensemble of bodies. These ``sub-bodies'' could represent hydrodynamical gas particles, or pebbles part of a larger asteroid \citep[e.g.][]{Quillen_etal_2016}. The net deformation is then calculated directly from the sum of their mutual forces, together with the external force from the other bodies. Alternatively, the internal structure can be approximated by a core and an envelope, with an associated spectrum of natural oscillation modes \citep[e.g.][]{Ogilvie_2009}. Particularly interesting is the excitation of discrete oscillation frequencies by a second orbiting body. This might leave an imprint in the gravitational wave signal of double white dwarfs, as well as in the flux of heartbeat stars. 

From the point of view of modelling $N$-body systems with $N \gg 3$ and/or long-term integrations, it is preferable to adopt fast and parametrized tidal models, which on average produce the correct dynamical evolution. A commonly adopted model is the linear tide model \citep[e.g.][]{Mignard_1979, Hut_1981}. Here, the bodies are assumed to take on their tidally deformed equilibrium shapes, but with a very small tidal time lag.  
The current-day satellite systems on near-circular orbits are an example in which linear tides are sufficient to describe their short-term evolution \citep[e.g.][]{Cuk_etal_2020}. On the other hand, for highly eccentric orbits, the linear approximation might not hold during pericenter passages, i.e. the tidal time lag might exceed the orbital and spin time scales. In this inverse regime, a non-linear or fast tide model is required \citep[e.g.][]{Correia_etal_2014}. Transitions from one tidal regime to another are expected to occur frequently in compact and chaotic multi-body systems. This motivates a tidal model which self-consistently handles the complete spectrum from linear to non-linear tides.

A more realistic description of the tidal response is to assume that the body always tends to creep towards the equilibrium by the 
action of the gravitational forces acting on it. The creep tide theory, or simply the Creep model, describes the response to tidal stress by supposing a linear creep equation for the shape evolution of the body \citep{Ferraz-Mello_2013}. For low-frequency perturbations, the Creep model is consistent with the linear tide model. However, the Creep model exhibits a peak dissipation when the tidal forcing period resonates with the bodies' fluid relaxation time. Further into the non-linear regime, tidal dissipation reduces again. Although there is a characteristic tidal frequency set by the fluid relaxation time, we note that this is different than the discrete oscillation frequencies considered by dynamic tide models. 

The Creep model also exhibits the same tidal dissipation as the visco-elastic Maxwell model, which is particularly well suited for reproducing the long-term deformation of the planets \citep[e.g.][]{Correia_etal_2014}. Therefore, it can be used to model the tidal evolution of these bodies more accurately than the linear model. For instance, it can handle capture in asynchronous spin-orbit resonances naturally \citep[e.g.][]{Valente_Correia_2022}. For stars and gaseous planets, the fluid relaxation time is expected to be several orders of magnitude smaller than for rocky planets \citep[see][Table~1]{Ferraz-Mello_2013}, and so the Creep model usually remains in the linear regime. However, during the early stages of the formation of hot-Jupiters and close binary stars, the orbits are expected to be very eccentric. As a result, the short-period passage at the pericentre, where tides are stronger, may require the use of the Creep model.
Furthermore, despite dynamic tides being a more accurate description of the dissipation within this class of bodies \citep[see][Fig.~2]{Ogilvie_2009}, the envelope of the multiple frequency perturbations, i.e. the averaged trend in the dissipation spectra, is consistent with a Creep model.

\texttt{TIDYMESS} is designed for long-term studies, where the orbits can present chaotic evolution and change from very eccentric to circular.
Therefore, the Creep model is better suited than the linear model, as it can handle different dissipation regimes.
Although the Creep model is still a simple tidal model characterised by a single parameter, the relaxation time $\taub$, for long-term studies we do not focus on subtle details such as a given dissipation peak at a specific forcing frequency. 
On the contrary, we mostly care about the overall picture. For short and long-term perturbations, we know that the dissipation has to vanish, but at some point in the middle there is a maximum dissipation, which can be calibrated by means of $\taub$.

\subsection{Outline and novelty}

We present the new $N$-body code \texttt{TIDYMESS}\footnote{\texttt{TIDYMESS} is available by request from the authors. It will be made available through an online repository in the near future.}, which stands for the ``TIdal DYnamics of Multi-body ExtraSolar Systems''. This code covers a niche regime of parameter space where orbits are chaotic and tides can become non-linear. This is accomplished by implementing the Creep tidal model, with a fourth-order direct integration method. The interplay between orbital, spin and shape evolution is calculated self-consistently. Arbitrary number of bodies and configurations can be simulated within a Cartesian inertial frame, including full information on the spin vectors and inertia tensors of the bodies. The main numerical novelty of our method is the ability to integrate the deformation over an orbital time step, irrespective of its fluid relaxation time or spin period. Our code complements the repertoire of tidal $N$-body codes in the literature \citep[e.g.][]{Bolmont_etal_2015, Tamayo_2020, Trani_2022}, and facilitates comparison and scrutiny of tidal models, which can be tested against current and upcoming observations of compact, multi-body systems. 

We present the theoretical framework of the tidal model in Sec.~\ref{sec:model}. The numerical implementation is described in Sec.~\ref{sec:numerics}. A gallery of demonstrations of \texttt{TIDYMESS} is presented in Sec.~\ref{sec:results}, including planetary and stellar applications. The main take away messages are gathered in Sec.~\ref{sec:conclusions}.   


\section{Model}\label{sec:model}

In this section we derive a very general model that is suited to study a system of $N$ bodies.
Our model is valid in 3D for the orbital planes and individual spins. 
We use Cartesian inertial coordinates.

\llabel{secmodel}

\subsection{Tidal model}
 
\subsubsection{Gravitational potential of an extended body}

We consider an extended body of mass $\m$, and choose as reference the Cartesian inertial frame ($\vi,\vj,\vk$).
In this frame, the angular velocity and rotational angular momentum vectors of the body are given by $\vw = (\w_\ia, \w_\jb, \w_\kc)$ and $\vL = (L_\ia, L_\jb, L_\kc)$, respectively, which are related through the inertia tensor $\TI$ as
\be
\vL = \TI \cdot \vw  \quad \Leftrightarrow \quad \vw = \TI^{-1} \cdot \vL 
\ ,
 \llabel{151019b}
\ee
where
\be
\TI = 
\left[\begin{array}{rrr} 
I_{11}&  I_{12}& I_{13} \crm
I_{12}&  I_{22}& I_{23} \crm
I_{13}&  I_{23}& I_{33} 
\end{array}\right] \ ,
\label{121026c}
\ee
\be\TI^{-1}  = \frac{1}{\Delta \TI}
\left[\begin{array}{ccc} 
I_{22} I_{33} - I_{23}^2   &  I_{13} I_{23} - I_{12} I_{33}   & I_{12} I_{23} - I_{22} I_{13} \crm
I_{13} I_{23} - I_{12} I_{33}   & I_{11} I_{33} - I_{13}^2    & I_{12} I_{13} - I_{11} I_{23} \crm
I_{12} I_{23} - I_{22} I_{13}   & I_{12} I_{13} - I_{11} I_{23}   & I_{11} I_{22} - I_{12}^2
\end{array}\right] \ ,
\label{151028g}
\ee
with
\be
\Delta \TI = I_{11} I_{22} I_{33} + 2 I_{12} I_{13} I_{23} - I_{11} I_{23}^2 - I_{22} I_{13}^2 - I_{33} I_{12}^2 
\ . \label{151028h}
\ee

The gravitational potential of the body at a generic position $\vr$ from its center-of-mass is given by \citep[e.g.][]{Goldstein_1950}
\be 
V (\vr) = - \frac{G \m}{r} + \frac{3 G}{2 r^3} \left[ \ur \cdot \TI \cdot \ur - \frac{1}{3} \mathrm{tr}(\TI) \right] \ , \llabel{121026b}
\ee
where $G$ is the gravitational constant, $\ur = \vr / r = (\xii, \xj, \xk)$ is the unit vector, and $\mathrm{tr}(\TI) = I_{11} + I_{22} + I_{33} $.
We neglect terms in $(R/r)^3$, where $R$ is the average radius of the body (quadrupolar approximation).
Adopting the Lagrange polynomial $P_2(x)=(3x^2-1)/2$, we can rewrite the previous potential as
\begin{eqnarray}
V (\vr) = - \frac{G \m}{r} 
+ \frac{G}{r^3}  \Big[ \big(I_{22}-I_{11}\big) P_2 (\xj) + \big(I_{33}-I_{11}\big) P_2 (\xk) && \nonumber \\ 
+ \, 3 \big( I_{12} \xii \xj + I_{13} \xii \xk + I_{23} \xj \xk \big) \!\!\!\!\!\! &  \Big] & \!\!\!\!\!\!
\ . \llabel{151028a}
\end{eqnarray}

\subsubsection{Deformation of the extended body}

The mass distribution inside an extended body, characterised by its inertia tensor, is a result of the forces acting on it, that is, self gravity and the body's response to any perturbing potential.
Here, we consider that the extended body deforms under the action of the centrifugal and tidal potentials.

For a body with rotation rate $\vw$, a mass element $d m$ at a location $\vr'$  is subject to a centrifugal potential \citep[e.g.][]{Goldstein_1950}
\be
V_c (\vr') = - \frac{1}{2} (\vw \times \vr')^2 = - \frac{1}{2} \w^2 r'^2 + \frac{1}{2} \left(\vw \cdot \vr'\right)^2 \ . \llabel{121026d}
\ee

Assuming also that the body moves in the field of a point-mass $\M$ located at $\vr$, a mass element $d m$ at a location $\vr'$  is subject to a tidal potential \citep[e.g.][]{Lambeck_1980}
\be
V_t (\vr') = -\frac{G \M}{r} \left(\frac{r'}{r}\right)^2 P_2 (\ur' \cdot \ur) \ , \llabel{121026e}
\ee
where we have neglected again terms higher than $(r'/r)^3$ . 

The resulting perturbing potential is then given by
\be
V_p (\vr') = V_c (\vr') + V_t (\vr')  \llabel{121026f} \ ,
\ee
that can also be rearranged as
\be
V_p (\vr') = -  \frac{r'^2 \w^2}{3} + \frac{3 G}{2 r'^3} \left[ \ur' \cdot \TI_p' \cdot \ur' -\frac{1}{3} \mathrm{tr}(\TI_p) \right]  \llabel{121030a} \ ,
\ee
with
\be
\frac{\TI_p'}{\m r'^2} = \frac{r'^3}{3 G \m} \vw \, \vw^T
-  \frac{\M}{\m} \left( \frac{r'}{r} \right)^3 \ur \, \ur^T  \llabel{121030b0} \ ,
\ee
where $^T$ denotes the transpose.
Alternatively, we can write
\be
V_p (\vr') = -  \frac{r'^2 \w^2}{3} + \frac{3 G}{2 r'^3} \left[ \ur' \cdot \TI_p \cdot \ur'  \right]  \llabel{171110z} \ ,
\ee
with
\be
\TI_p = \TI_p' -\frac{1}{3} \mathrm{tr}(\TI_p') \mathbb{I} 
\llabel{171110y} \ ,
\ee
where $\mathbb{I}$ is the identity tensor.
Thus, on the body's surface, $\vR$, 
the non-spherical contribution of the perturbing potential is given by
\be
V_p (\vR) = \frac{3 G}{2 R^3} \left[ \hat \vR \cdot \TI_p \cdot \hat \vR  \right] 
\ , \llabel{171110x}
\ee
where $R = |\vR|$.
For simplicity, when the surface of the body is nearly spherical, we can assume $R$ to be constant and equal to the average radius of the body.

A most convenient way of handling the deformation is through the Love number approach \citep[e.g.][]{Love_1927, Peltier_1974}.
As long as the distortions are small, we can simplify the problem by ignoring the small
interaction terms between the centrifugal and tidal potentials \citep{Zharkov_Trubitsyn_1978}.
The equilibrium potential $V_\eq$ is then obtained for static perturbations or instantaneous response simply as
\be
V_\eq (\vR) = \kf V_p (\vR) =  \frac{3 G}{2 R^3} (\hat \vR \cdot \TI_\eq \cdot \hat \vR)
\ , \llabel{130528z}
\ee
with (Eq.\,(\ref{171110y}))
\be
\frac{\TI_\eq}{\m R^2} =  \frac{\kf R^3}{3 G \m} \left( \vw \, \vw^T - \frac{\w^2}{3}  \mathbb{I} \right)  - \kf \frac{\M}{\m} \left( \frac{R}{r} \right)^3 \left( \ur \, \ur^T - \frac{1}{3}  \mathbb{I} \right)
 \llabel{121030b} \ ,
\ee
where $\kf$ is the fluid second Love number for potential.
By comparing expression (\ref{130528z}) with the gravitational potential (\ref{121026b}) at the body's surface, we see that $\TI_\eq$ corresponds to the non-spherical contribution to the inertia tensor.
The values for each coefficient of $\TI_\eq$ are
\be
\frac{I_{11}^\eq}{\m R^2} =  \frac{\kf R^3}{3 G \m} \left( \w_\ia^2 -\frac{\w^2}{3} \right) - \kf \frac{\M}{\m} \left(\frac{R}{r}\right)^3 \left( \xii^2 - \frac{1}{3} \right)
 \ , \llabel{151106a}
\ee
\be
\frac{I_{22}^\eq}{\m R^2} = \frac{\kf R^3}{3 G \m} \left( \w_\jb^2 -\frac{\w^2}{3} \right) - \kf \frac{\M}{\m} \left(\frac{R}{r}\right)^3 \left( \xj^2 - \frac{1}{3} \right)
\ , \llabel{151106b} 
\ee
\be
\frac{I_{33}^\eq}{\m R^2} = \frac{\kf R^3}{3 G \m} \left( \w_\kc^2 -\frac{\w^2}{3} \right) - \kf \frac{\M}{\m} \left(\frac{R}{r}\right)^3 \left( \xk^2 - \frac{1}{3} \right)
\ , \llabel{151106c} 
\ee
\be
\frac{I_{12}^\eq}{\m R^2} = \frac{\kf R^3}{3 G \m} \w_\ia \w_\jb - \kf \frac{\M}{\m} \left(\frac{R}{r}\right)^3 \xii \xj
\ , \llabel{151106d} 
\ee
\be
\frac{I_{13}^\eq}{\m R^2} = \frac{\kf R^3}{3 G \m} \w_\ia \w_\kc - \kf \frac{\M}{\m} \left(\frac{R}{r}\right)^3 \xii \xk
\ , \llabel{151106e}
\ee
\be
\frac{I_{23}^\eq}{\m R^2} = \frac{\kf R^3}{3 G \m} \w_\jb \w_\kc - \kf \frac{\M}{\m} \left(\frac{R}{r}\right)^3 \xj \xk
\ . \llabel{151106f} 
\ee

In general, the perturbing potential is not constant and the deformation of the body is not instantaneous.
However, for the Creep model, in the body frame $B$, the inertia tensor of the body $\TI^B$, and the instantaneous equilibrium inertia tensor $\TI^B_\eq$, can be related through a simple differential equation as \citep{Correia_etal_2014}:
\be
\TI^B + \taub \dot \TI^B =  \TI_0 + \TI^B_\eq 
\ , \llabel{130107p} 
\ee
where $\taub = 19 \nue / (2 g \rho R)$ is the fluid relaxation time, $\nue$ is the viscosity, 
$g$ is the surface gravity and 
$\rho$ is the mean density.
$\TI_0 = \xix \m R^2 \, \mathbb{I} $ corresponds to the spherical contribution to the inertia tensor of the body, where $\xix$ is the inner structure factor (we have $\xix = 2/5$ for an homogenous body).

We let $\SR$ be the rotation matrix that allows us to convert any vector $\vu_B$ in a frame attached to the body into the Cartesian inertial frame $\vu$, such that $\vu = \SR \, \vu_B$.
The time evolution of $\SR$ is given by 
\be
\dot \SR = \tvw \, \SR  \ , \quad \mathrm{and} \quad \dot \SR^T = - \SR^T \tvw  
\ , \llabel{160127b}
\ee
with 
\be
\tvw = \left[\begin{array}{ccc} 
0   &  - \w_\kc   & \w_\jb \crm
\w_\kc   & 0   & - \w_\ia \crm
- \w_\jb   & \w_\ia   & 0
\end{array}\right] 
\ . \llabel{160111b}
\ee
In the inertial frame, 
\be
\TI = \SR \TI^B \SR^T 
\ , \llabel{160127a}
\ee
and so we can rewrite expression (\ref{130107p}) as
\be
\TI + \taub \SR \dot \TI^B \SR^T = \TI_0 + \TI_\eq 
\ . \llabel{160111d} 
\ee
Now, differentiating equation (\ref{160127a}), and using the derivatives of $\SR$ given by (\ref{160127b}), we have
\be
\dot \TI = \tilde \vw \TI - \TI \tilde \vw + \SR \dot \TI^B \SR^T \ ,
\llabel{160111e}
\ee
or, using equation (\ref{160111d})
\be
\dot \TI  =  \tilde \vw \TI - \TI \tilde \vw + ( \TI_0 + \TI_\eq - \TI ) / \taub 
\ . \llabel{160111f} 
\ee

For incompressible bodies, $\mathrm{tr}(\TI)=\mathrm{tr}(\TI_0) = 3 \xix \m R^2$. 
We thus have, $I_{33} = 3 \xix \m R^2 - (I_{11} + I_{22}) $, 
that is, the inertia tensor has only five independent coefficients:
\be
\left\{\begin{array}{l} 
\dot \Iv_{11} =  2 \Iv_{13}\w_\jb - 2\Iv_{12}\w_\kc + (\xix \m R^2 + 
I_{11}^\eq-\Iv_{11})/ \tau \crm
\dot \Iv_{22} =  -2 \Iv_{23}\w_\ia + 2\Iv_{12}\w_\kc + (\xix \m R^2 + 
I_{22}^\eq-\Iv_{22})/ \tau \crm
\dot \Iv_{12} =  - \Iv_{13}\w_\ia + \Iv_{23}\w_\jb + (\Iv_{11}-\Iv_{22})\,\w_\kc + (I^\eq_{12}-\Iv_{12})/ \tau \crm
\dot \Iv_{13} =  \Iv_{12}\w_\ia - (2 \Iv_{11}+\Iv_{22})\,\w_\jb  -  \Iv_{23}\w_\kc + (I^\eq_{13}-\Iv_{13})/ \tau \crm
\dot \Iv_{23} =  (\Iv_{11}+2 \Iv_{22})\,\w_\ia  - \Iv_{12}\w_\jb +  \Iv_{13}\w_\kc + (I^\eq_{23}-\Iv_{23})/ \tau
\end{array} \right.  
\ . \label{171109b}
\ee

\subsubsection{Linear approximation}
\label{sec:linearapprox}

When the relaxation time, $\tau$, is much smaller than the orbital and spin evolution timescales, the integration time of the system is dominated by the deformation equations (\ref{160111f}).
However, for small $\tau$ values, we can assume that the deformations are linear and directly obtain an expression for $\TI$.
From expression (\ref{130107p}) we compute
\be
\TI^B (t) = \TI_0 + \frac{\ep^{-t/\tau}}{\tau}  \int_0^t \TI^B_\eq (x) \, \ep^{x/\tau} dx + {\cal C}_1 \, \ep^{-t/\tau}
\ , \llabel{200114a} 
\ee
where ${\cal C}_k$ is an integration constant.
Integrating by parts we get
\be
\int_0^t \TI^B_\eq (x) \, \ep^{x/\tau} dx = {\cal C}_2 + \tau \TI^B_\eq (t) \, \ep^{t/\tau} - \tau^2 \dot \TI^B_\eq (t) \, \ep^{t/\tau} + {\cal O} (\tau^3)
\ , \llabel{200114b} 
\ee
and thus,
\be
\TI^B (t) = \TI_0 + \TI^B_\eq (t) - \tau \dot \TI^B_\eq (t) + {\cal C}_0 \, \ep^{-t/\tau} + {\cal O} (\tau^2)
\ . \llabel{200114c} 
\ee
In the following, we neglect the term ${\cal C}_0 \, \ep^{-t/\tau}$ because it vanishes after some time.
Replacing in expression (\ref{160127a}) we get
\be
\TI = \TI_0 + \TI_\eq - \tau \, \SR \, \dot \TI^B_\eq \SR^T + {\cal O} (\tau^2)
\ , \llabel{200114d} 
\ee
or, similarly (Eq.\,(\ref{160111e}))
\be
\TI = \TI_0 + \TI_\eq + \tau \left(\tilde \vw \TI_\eq - \TI_\eq \tilde \vw - \dot \TI_\eq \right) + {\cal O} (\tau^2)
\ . \llabel{200114e} 
\ee

\subsection{$N$-body problems}

\subsubsection{Point-mass problem}

\llabel{pmp}

We now consider that the extended body orbits a point-mass $\M$ located at $\vr$.
The force between the two bodies is easily obtained from the potential energy of the system $ U (\vr) = \M V (\vr) $  (Eq.\,(\ref{151028a})) as 
\be
\vF = - \nabla U (\vr) =  \vf(\M, \m, \vr) + \vg(\M, \TI, \vr) + \vh(\M, \TI, \vr) 
\ , \llabel{170911d}
\ee
with
\be
\vf(\M, \m, \vr) = - \frac{G \M \m}{r^3} \vr \ ,  \llabel{170911a}
\ee
\begin{eqnarray}
\vg(\M, \TI, \vr) \!\!\!\!\! & = & \!\!\!\!\!  
\frac{15 G \M}{r^5}  \Big[ \frac{I_{22}-I_{11}}{2} \big(\xj^2 - \frac15 \big)  
+ \frac{I_{33}-I_{11}}{2} \big(\xk^2 - \frac15 \big) \nonumber \\ 
&& \!\!\!\!\! + I_{12} \xii \xj + I_{13} \xii \xk + I_{23} \xj \xk \Big] \vr 
\ , \llabel{170911b}
\end{eqnarray}
\begin{eqnarray}
\vh(\M, \TI, \vr) \!\!\!\!\! & = & \!\!\!\!\! 
 -  \frac{3 G \M}{r^4}  \Big[ \big(I_{22}-I_{11}\big) \xj \vj + \big(I_{33}-I_{11}\big) \xk \vk \nonumber \\ 
 && \!\!\!\!\! + I_{12} (\xii \vj + \xj \vi)  + I_{13} (\xii \vk + \xk \vi) + I_{23} (\xj \vk + \xk \vj) \Big]
\ . \llabel{170911c}
\end{eqnarray}
We thus obtain for the orbital evolution of the system
\be
\ddvr =  \vv{F} / \beta \ ,  \llabel{151028c}
\ee
where $\beta = \M \m/ (\M + \m)$ is the reduced mass.
The spin evolution of the extended body can also be obtained from the force, by computing the gravitational torque. In the inertial frame we have: 
\be
\dvL = \vT (\M, \TI, \vr) = - \vr \times \vv{F} = - \vr \times \vh 
\ , \llabel{150626a}
\ee
that is,
\begin{eqnarray}
\vT (\M, \TI, \vr) = \frac{3 G \M}{r^3} \ur \times \Big[ \big(I_{22}-I_{11}\big) \xj \vj + \big(I_{33}-I_{11}\big) \xk \vk && \nonumber \\ 
+ I_{12} (\xii \vj + \xj \vi) + I_{13} (\xii \vk + \xk \vi) + I_{23} (\xj \vk + \xk \vj) \!\!\!\!\!\! &  \Big] & \!\!\!\!\!\!
\ , \llabel{151028d}
\end{eqnarray}
or
\be
\vT =  \frac{3 G \M}{r^3}
\left[\begin{array}{c}  
\big(I_{33}-I_{22}\big) \xj \xk - I_{12} \xii \xk + I_{13} \xii \xj  + I_{23} (\xj^2 - \xk^2)  \crm 
\big(I_{11}-I_{33}\big) \xii \xk + I_{12} \xj \xk + I_{13} (\xk^2 - \xii^2) - I_{23} \xii \xj  \crm 
\big(I_{22}-I_{11}\big) \xii \xj + I_{12} (\xii^2 - \xj^2) - I_{13} \xj \xk + I_{23} \xii \xk
\end{array}\right] 
\ . \llabel{151028e}
\ee

To solve the spin-orbit motion, we need to integrate equations (\ref{160111f}), (\ref{151028c}) and (\ref{150626a}).


\subsubsection{Two-body problem}

Consider now that two extended bodies with masses $\m_0$ and $\m_1$, and inertia tensors $\TI_0$ and $\TI_1$, respectively, orbit around each other at a distance $\vr$ from their centers-of-mass.
The total potential energy can be written from expression (\ref{121026b}) as
\be 
U = - \frac{G \m_0 \m_1}{r} + \frac{3 G}{2 r^3} \left[ \ur \cdot \TJ \cdot \ur - \frac{1}{3} \mathrm{tr}(\TJ) \right] \ , \llabel{170911e}
\ee
with $\TJ = \m_0 \TI_1 + \m_1 \TI_0$.
This potential is very similar to the previous point-mass problem and the equations of motion are simply
\be
\ddvr  = \vF_{01} (\vr) / \beta_{01} \ , \llabel{170911f}
\ee
\be
\dvL_0 = \vT_{01} (\vr) \ , \quad  \dvL_1 = \vT_{10} (\vr)
\ , \llabel{170911g}
\ee
where
\be
\begin{split}
\vF_{kl} (\vr)  =
\vf(\m_k, \m_l, \vr) + \vg(\m_k, \TI_l, \vr) + \vg(\m_l, \TI_k, \vr) \\ 
 + \vh(\m_k, \TI_l, \vr) + \vh(\m_l, \TI_k, \vr)
\ , \llabel{170911f}
\end{split}
\ee
\be
 \vT_{kl} (\vr) =  \vT (\m_l, \TI_k, \vr)
\ , \llabel{170911g}
\ee
$\beta_{kl} = \m_k \m_l / (\m_k + \m_l)$, $\vL_k = \TI_k \vw_k $ is the rotational angular momentum vector of the body with mass $\m_k$, and $\vw_k$ its angular velocity.

\subsubsection{$N$-body problem}
\llabel{nbody}

The previous equations can be easily generalised to the motion of several bodies.
For a system of $N+1$ extended bodies, with masses $\m_k$ and inertia tensors $\TI_k$ ($k = 0,1,...,N$), the total potential energy can be written from expression (\ref{170911e}) as
\be 
U = \sum_{k=0}^N \sum_{l>k}^N \left( - \frac{G \m_k \m_l}{\rkl} + \frac{3 G}{2 \rkl^3} \left[ \urkl \cdot \TJ_{kl} \cdot \urkl - \frac{1}{3} \mathrm{tr}(\TJ_{kl}) \right] \right) \ , \llabel{170911i}
\ee
with  $\TJ_{kl} = \m_k \TI_l + \m_l \TI_k$, and $\vrkl = \vR_k - \vR_l$, where $\vR_k$ is the position of the center-of-mass of the body $k$ in the inertial frame.
The equations of motion for each body in the inertial frame are thus
\be
\ddvR_k = \vv{a}_k = \frac{1}{\m_k} \sum_{l\ne k} \vF_{kl} (\vrkl) \llabel{170911j} \ ,
\ee
\be
\dvL_k = \sum_{l\ne k} \vT_{kl} (\vrkl) \ ,  \llabel{170911k}
\ee
where $\vF_{kl} (\vrkl)$ and $\vT_{kl} (\vrkl)$ are given by expressions (\ref{170911f}) and  (\ref{170911g}), respectively.
As for the two-body problem, we can also express the motion in the relative frame, for instance, with respect to the body with mass $\m_0$.
We let $\vr_k = \vr_{k0} =  \vR_k - \vR_0 \, (k=1,...,N)$.
Thus, 
\be
\ddvr_k  =  \frac{1}{\beta_{0k}} \vF_{0k}(\vr_k) + \sum_{l\ne k} \left[ \frac{1}{\m_k} \vF_{kl} (\vr_k-\vr_l) + \frac{1}{\m_0} \vF_{0l} (\vr_l) \llabel{170911m} \right] \ .
\ee
Note also that for the equilibrium inertia tensor (Eq.\,(\ref{121030b})), we now need to take into account the tidal perturbations from all bodies in the system:
\be
\TI_{\eq, k} = \frac{\kf R_k^5}{3 G} \left( \vw_k \, \vw_k^T - \frac{\w_k^2}{3}  \mathbb{I} \right)  - \sum_{l\ne k} \kf \m_l R_k^2 \left( \frac{R_k}{r_{kl}} \right)^3 \left( \ur_{kl} \, \ur_{kl}^T - \frac{1}{3}  \mathbb{I} \right) \llabel{220830a} \ ,
\ee
where $\kf$ also pertains to the body with index $k$.

\subsection{Other physical phenomena}

\subsubsection{General relativity corrections}
\label{section:grc}

Relativistic effects become important in close binary systems, through apsidal precession and gravitational wave emission leading to orbital shrinkage. Relativistic perturbations also play a role in large stellar systems by affecting the global chaotic properties, e.g. the Lyapunov time scale \citep{spz22}. Furthermore, interesting phenomena are expected to result from the interplay between relativity and tides in compact multiple systems. 

A common method for implementing relativistic effects in $N$-body codes is by adding Post-Newtonian (PN) acceleration terms. The force is expanded in a series based on the small ratio of velocity divided by the speed of light. We will include the pairwise, conservative 1PN and 2PN terms, as well as the first dissipative term, 2.5PN. 
The Post-Newtonian force experienced by body $k$ needed to correct expression (\ref{170911j}) is given by
\begin{equation}
    \vv{F}_{\rm{PN, k}} = m_k \left( \vv{a}_{\rm{1PN,k}} + \vv{ a}_{\rm{2PN,k}} + \vv{a}_{\rm{2.5PN,k}} \right).     
\end{equation}

\noindent The 1PN acceleration term is given by
\begin{equation}
\vv{a}_{\rm{1PN,k}} = \frac{1}{c^2} \sum_{l \neq k} \frac{G m_l}{r_{kl}^2} \left( B_{1,kl}\, \vv{\hat{r}}_{kl} + C_{1,kl} \, \vv{v}_{kl} \right),     
\end{equation}

\noindent with $\vv{v}_{kl} = \vv{\dot{r}}_{kl}$, and $c$ the speed of light. The coefficients are
\begin{equation}
    B_{1,kl} = 5 \frac{G m_k}{r_{kl}} + 4 \frac{G m_l}{r_{kl}} - v_k^2 - 2v_l^2 + 4\vv{v}_k \cdot \vv{v}_l + \frac{3}{2} \left( \vv{\hat{r}}_{kl} \cdot \vv{v}_l \right)^2,
\end{equation}
\begin{equation}
    C_{1,kl} = 4 \vv{\hat{r}}_{kl} \cdot \vv{v}_k - 3 \vv{\hat{r}}_{kl} \cdot \vv{v}_l.
\end{equation}

\noindent The 2PN acceleration term is given by
\begin{equation}
\vv{a}_{\rm{2PN,k}} = \frac{1}{c^4} \sum_{l \neq k} \frac{G m_l}{r_{kl}^2} \left( B_{2,kl}\, \vv{\hat{r}}_{kl} + C_{2,kl} \, \vv{v}_{kl} \right) + D_{2,kl} \vv{\hat{r}}_{kl},     
\end{equation}
\noindent with
\begin{eqnarray}
    B_{2,kl} = -2 v_l^4 + 4 v_l^2 \vv{v}_{k} \cdot \vv{v}_l - 2 \left( \vv{v}_{k} \cdot \vv{v}_l \right)^2 - \frac{15}{8} \left( \vv{\hat{r}}_{kl} \cdot \vv{v}_l \right)^4 \nonumber \\
    + \left( \frac{3}{2} v_k^2 + \frac{9}{2} v_l^2  - 6 \vv{v}_k \cdot \vv{v}_l \right) \left( \vv{\hat{r}}_{kl} \cdot \vv{v}_l \right)^2 \nonumber \\
    + \frac{G m_l}{r_{kl}} \left( 4 v_l^2 - 8 \vv{v}_k \cdot \vv{v}_l + 2 \left( \vv{\hat{r}}_{kl} \cdot \vv{v}_k \right)^2 \right. \nonumber \\
    \left. - 4 \left( \vv{\hat{r}}_{kl}  \cdot \vv{v}_k \right)\left( \vv{\hat{r}}_{kl} \cdot \vv{v}_l \right) - 6 \left( \vv{\hat{r}}_{kl} \cdot \vv{v}_l \right)^2 \right) \nonumber \\
    + \frac{G m_k}{r_{kl}} \left( -\frac{15}{4} v_k^2 + \frac{5}{4} v_l^2 - \frac{5}{2} \vv{v}_k \cdot \vv{v}_l + \frac{39}{2} \left( \vv{\hat{r}}_{kl} \cdot \vv{v}_k \right)^2 \right. \nonumber \\ 
    \left. - 39 \left( \vv{\hat{r}}_{kl}  \cdot \vv{v}_k \right)\left( \vv{\hat{r}}_{kl} \cdot \vv{v}_l \right) + \frac{17}{2} \left( \vv{\hat{r}}_{kl} \cdot \vv{v}_l \right)^2 \right)
\end{eqnarray}
\begin{eqnarray}
    C_{2,kl} = v_k^2 \vv{\hat{r}}_{kl} \cdot \vv{v}_l + 4 v_l^2 \vv{\hat{r}}_{kl} \cdot \vv{v}_k - 5 v_l^2 \vv{\hat{r}}_{kl} \cdot \vv{v}_l - 4 \left( \vv{v}_k \cdot \vv{v}_l \right) \left( \vv{\hat{r}}_{kl} \cdot \vv{v}_k \right) \nonumber \\
    + 4 \left( \vv{v}_k \cdot \vv{v}_l \right) \left( \vv{\hat{r}}_{kl} \cdot \vv{v}_l \right) - 6 \left( \vv{\hat{r}}_{kl} \cdot \vv{v}_k \right) \left( \vv{\hat{r}}_{kl} \cdot \vv{v}_l \right)^2 + \frac{9}{2} \left( \vv{\hat{r}}_{kl} \cdot \vv{v}_l \right)^3 \nonumber \\
    + \frac{G m_k}{r_{kl}} \left( -\frac{63}{4} \vv{\hat{r}}_{kl} \cdot \vv{v}_k + \frac{55}{4} \vv{\hat{r}}_{kl} \cdot \vv{v}_l \right) \nonumber 
    + \frac{G m_l}{r_{kl}} \left( -2 \vv{\hat{r}}_{kl} \cdot \vv{v}_k - 2 \vv{\hat{r}}_{kl} \cdot \vv{v}_l \right)
\end{eqnarray}
\begin{eqnarray}
    D_{2,kl} = \frac{G^3 m_l}{r_{kl}^4}\left( -\frac{57}{4} m_k^2 - 9 m_l^2 - \frac{69}{2} m_k m_l \right). 
\end{eqnarray}

\noindent The 2.5PN acceleration term is given by
\begin{equation}
\vv{a}_{\rm{2.5PN,k}} = \frac{1}{c^5} \sum_{l \neq k} \frac{4}{5}\frac{G^2 m_k m_l}{r_{kl}^3} \left( B_{25,kl}\, \vv{\hat{r}}_{kl} + C_{25,kl} \, \vv{v}_{kl} \right),     
\end{equation}

\noindent with
\begin{eqnarray}
B_{25,kl} = \left( 3 v_{kl}^2 - 6 \frac{G m_k}{r_{kl}} + \frac{52}{3} \frac{G m_l}{r_{kl}} \right) \left( \vv{\hat{r}}_{kl} \cdot \vv{v}_k - \vv{\hat{r}}_{kl} \cdot \vv{v}_l \right),
\end{eqnarray}
\begin{eqnarray}
C_{25,kl} = -v_{kl}^2 + 2\frac{G m_k}{r_{kl}} - 8\frac{G m_l}{r_{kl}}.
\end{eqnarray}



\subsubsection{Magnetic braking}
\label{section:mbe}

Low-mass stars lose angular momentum due to spin-down magnetic braking through the action of magnetized stellar winds  \citep{Skumanich_1972, Kawaler_1988}.
Cool stars are thought to be more efficient in generating magnetic winds and thus lose angular momentum more rapidly than hot stars \citep[e.g.][]{Weber_Davis_1967}.
An empirical relation can be derived from the observations, where $\dot \w \propto - \w^3$.
For Sun-like stars we can use this simple law to compute the contribution of magnetic braking to the angular momentum variation of each star in the model:
\be
\dot \vL_k = - \alpha_k \, \w_k^2 \, \vL_k \ , \llabel{190129z}
\ee
where $\alpha_k$ is a proportionality constant determined from the observations.
For G and K-dwarf stars, we estimate $\alpha_k \approx 1.5 \times 10^{-14}$~year \citep{Barker_Ogilvie_2009}.
For F-dwarfs the magnetic braking is less efficient and may be only 10\% of the previous value.
Therefore, in \texttt{TIDYMESS} it is possible to redefine the value of $\alpha_k$ for each star.

\subsubsection{Mergers and collisions}
\label{section:col}

Mergers and collisions are natural outcomes in compact, multi-body systems \citep[e.g.][]{Vergara21, Toonen22}. For example, in resonant multi-planet systems, planet-planet interactions can lead to one planet being scattered towards the host star. There is a thin line between either a collision or tidal capture. In the latter case, the subsequent tidal evolution might lead to the formation of a hot Jupiter (see Sec.~\ref{sec:ex5}). We will therefore include stopping conditions for collisions, based on the criterion of overlapping radii.

Optionally, the simulations can be continued by replacing the collisional components by an approximate collision product. However, this requires further details on the internal structure properties and corresponding tidal response parameters as a function of mass and impact properties. \texttt{TIDYMESS} implements a generalisation of the sticky sphere model, which assumes the conservation of mass, center of mass, linear momentum and total angular momentum (orbit and spin). Internal properties are adopted from the most massive collision component, e.g. the radius is determined by preserving the average density. Alternatively, the code can be halted when a collision is detected, and a collision product can be generated externally, e.g. using the \texttt{AMUSE} framework \citep{spz_2018} and specialised codes for collisions \citep[e.g.][]{gaburov_2018}. 


\section{Numerical implementation}\label{sec:numerics}

In this section we provide details on the numerical implementation of \texttt{TIDYMESS}, based on the physical model given in the previous section. First, we discuss the tidal models implemented in \texttt{TIDYMESS}, followed by our general and optimized approximation of the Creep model integration method. Then, we describe the integration maps, time step functions and the user interface.\footnote{Further code tutorials and examples are provided with the source code. }

\subsection{Tidal models}

\texttt{TIDYMESS} includes five tidal models/implementations, which are listed in Tab.~\ref{tab:tidal_models}.
The most general and accurate model is the tidal Creep model with \texttt{direct} integration of the deformation equations (Eq.\,(\ref{160111f})). This model is best used in the regime where the orbital time scales are of the same order or shorter than the spin and fluid relaxation time scales. For fast spinning bodies or very short fluid relaxation times, this method becomes excessively slow (see example in Sect.~\ref{sec:ex1}). This is because the integration step is required to be the minimum of all orbital periods, spin periods and fluid relaxation times. Therefore, we also implement a simplified Creep tidal model based on the algorithm described in Sec.~\ref{sec:new_method}, which we name \texttt{tidymess}. Using this model, the performance in the linear and non-linear tidal regimes is the same. A (small) approximation is introduced however, by the splitting of the tidal and rotational deformation parts (see example in Sect.~\ref{sec:ex1}). The third tidal model is the \texttt{linear} tide model (Eq.\,(\ref{200114e})). If the fluid relaxation time is always much shorter than the orbital and spin time scales in the system, then the Creep model reduces to the linear tide model. In this regime however, the linear tide model has a better performance. In the asymptotic limit of $\tau \rightarrow 0$, we obtain the \texttt{conservative} tide model, which assumes the instantaneous equilibrium shape for the deformations (Eq.\,(\ref{220830a})). Both angular momentum and energy are conserved in this model, and therefore also allows for backward integration. For completeness, we also implement a model for perfectly rigid spheres, which is the regime of $\kf \rightarrow 0$, i.e., the pure $N$-body point-mass problem.

\begin{table}
\centering
\begin{tabular}{| l c c |} 
 \hline
Tidal model  & Attributes & Notes \\
 \hline\hline
0. No tides & $m, \vv{r}, \vv{v}$ & Pure $N$-body code \\
1. \texttt{Conservative} tides & $R, \xi, \TI, \vw, \vL, \kf$ & Equilibrium shape \\
2. \texttt{Linear} tides & $\tau$ & Small tidal time lag \\
3. Creep tides \texttt{direct} & & Arbitrary time lag \\
4. Creep tides \texttt{tidymess} & & Orbital time steps only \\ 
 \hline
\end{tabular}
\caption{Tidal models implemented in \texttt{TIDYMESS}. Each model inherets the particle attributes from the models above: $m$ = mass, $\vv{r}$ = position vector, $\vv{v}$ = velocity vector, $R$ = mean radius, $\xi$ = moment of inertia factor, $\TI$ = inertia tensor, $\vw$ = spin frequency vector, $\vL$ = angular momentum vector, $\kf$ = fluid Love number, and $\tau$ = fluid relaxation time. }
\label{tab:tidal_models}
\end{table}

\subsection{Creep tides \texttt{tidymess}}\label{sec:new_method}


Analogous to the linear approximation (sect.~\ref{sec:linearapprox}), we start with the general solution to Eq.\,(\ref{130107p}): 

\begin{equation}
\TI^B\left( \Delta t \right) = \TI_0 + (\TI^B - \TI_0) \, {\ep}^{-{\Delta t}/{\tau}} + \frac{\ep^{-{\Delta t}/{\tau}}}{\tau} \int_0^{\Delta t} \TI_\eq^B\left( x \right) \ep^{ {x}/{\tau} } dx.
\end{equation}

\noindent Here $\Delta t$ is a timestep into the future, and for the integral we set the current time to $t=0$.
Since $\TI_0$ is constant, we define for simplicity $\TIv$ as the non spherical part of the inertia tensor
\be
\TIv = \TI - \TI_0 
\quad \Rightarrow \quad
\TI = \TI_0 + \TIv
\ .
\ee
Using integration by parts, 
we can write
\begin{equation}
\begin{split}
\TIv^B\left( \Delta t \right) = \TI_\eq^B\left( \Delta t \right) - \left( \TI_\eq^B - \TIv^B \right) \ep^{ -{\Delta t}/{\tau} } - \ep^{ -{\Delta t}/{\tau} } \int_0^{\Delta t} \dot{\TI}_\eq^B\left( x \right) \ep^{ {x}/{\tau} } dx.
\end{split}
\end{equation}

\noindent We perform another integration by parts, 
so that we can write
\begin{equation}
\begin{split}
\TIv^B\left( \Delta t \right) = \TI_\eq^B\left( \Delta t \right) - \tau \dot{\TI}_\eq^B\left( \Delta t \right) - \left(\TI_\eq^B - \tau\dot{\TI}_\eq^B - \TIv^B \right) \ep^{ -{\Delta t}/{\tau} } \\ 
 + \tau \ep^{ -{\Delta t}/{\tau} } \int_0^{\Delta t} \ddot{\TI}_\eq^B\left( x \right) \ep^{ {x}/{\tau} } dx.
\end{split}
\label{eq:Ifull}
\end{equation}

\noindent We could continue integrating by parts and add terms of increasing order in $\tau$. However, when $\tau \ll \Delta t$, the higher order terms will be very small, and we can therefore ignore terms of order $\tau^2$ and higher (linear approximation). If $\tau \gg \Delta t$ then we cannot ignore the higher order terms. But in that case, $\Delta t$ can be considered very small, which allows us to estimate the final integral in Eq.\,(\ref{eq:Ifull}) as

\begin{equation}
\tau \ep^{ -{\Delta t}/{\tau} } \int_0^{\Delta t} \ddot{\TI}_\eq^B\left( x \right) \ep^{ {x}/{\tau} } dx \approx \tau \ep^{ -{\Delta t}/{\tau} } \ddot{\TI}_\eq^B \Delta t \approx \tau \ep^{ -{\Delta t}/{\tau} } \left( \dot{\TI}_\eq^B\left( \Delta t\right) - \dot{\TI}_\eq^B \right) ,
\end{equation}

\noindent where we have estimated that $\ddot{\TI}_\eq^B \approx \left( \dot{\TI}_\eq^B\left(\Delta t \right) - \dot{\TI}_\eq^B \right)\,/\,\Delta t$.
The full expression for the inertia tensor in the body frame then becomes
\begin{equation}
\TIv^B\left( \Delta t \right) = \TI_\eq^B\left( \Delta t \right)
+ \left( \TIv^B - \TI_\eq^B \right) \ep^{ -{\Delta t}/{\tau}}
- \tau \dot{\TI}_\eq^B\left( \Delta t \right) \left( 1 - \ep^{ -{\Delta t}/{\tau} } \right).
\label{eq:Ifinal}
\end{equation}

\noindent The derivative of the inertia tensor is further approximated as $\dot{\TI}_\eq^B\left( \Delta t \right) \approx \left( \TI_\eq^B\left(\Delta t \right) - \TI_\eq^B \right)\,/\,\Delta t$, resulting in
\begin{equation}
\begin{split}
\TIv^B\left( \Delta t \right) = \TI_\eq^B\left( \Delta t \right)
+ \left( \TIv^B - \TI_\eq^B \right) \ep^{ -{\Delta t}/{\tau}} \\
- \frac{\tau}{\Delta t} \left( \TI_\eq^B\left( \Delta t\right) - \TI_\eq^B \right) \left( 1 - \ep^{ -{\Delta t}/{\tau} } \right).
\label{eq:Ifinal2}
\end{split}
\end{equation}

This equation is general and a good approximation for every value of the fraction $\Delta t/\tau$. To demonstrate this, we give the simplifying expressions in the different asymptotic regimes: 
\begin{itemize}
\item $\TIv^B\left( \Delta t \right) = \TI_v^B + (\TI_\eq^B-\TIv^B) \, {\Delta t}/{\tau}  \ , \quad $ if $\,{\Delta t}/{\tau} \ll 1$ \ ,  \\
\item $\TIv^B\left( \Delta t \right) = \TI_\eq^B\left( \Delta t \right) - \tau \dot{\TI}_\eq^B\left( \Delta t \right) \ , \quad $ if $\,{\Delta t}/{\tau} \gg 1$ \ . 
\end{itemize}

\noindent The expression above is for the inertia tensor in the body frame, i.e., it is rotating with the body. In order to get the inertia tensor in the inertial frame, we need to perform a rotation from the body to the inertial frame (see Eq.\,(\ref{160127a})). Furthermore, the change of the inertia tensor occurs on two separate time scales: the spin time scale which is proportional to $\w^{-1}$, and the fluid relaxation time scale, $\tau$ (see Eq.\,(\ref{160111f})). Since these time scales can be very different, it is natural to split up the differential equation of the inertia tensor into two parts, in a manner that is often done in Hamiltonian splittings: 

\begin{equation}
\dot{\TI} = (\TI_\eq - \TIv) / \tau \,\,\,\,\rm{(Deformation\,part)}.
\end{equation}

\begin{equation}
\dot{\TI} = \tilde{\vw} \TI - \TI \tilde{\vw}\,\,\,\,\rm{(Rotational\,part)},
\end{equation}

\noindent Each of these parts separately can be solved accurately and without limits on the time step size. The solution to the deformation part is given in Eq.\,(\ref{eq:Ifinal2}) without the body frame superscripts. The rotational part can be problematic if the body is spinning very rapidly, e.g. when $\w \Delta t > 1$ or $\w \tau > 1$, then we cannot simply do $\TI (\Delta t) = \TI + \dot{\TI}\Delta t$. This can be solved accurately using a (quaternion) rotation operator. 
However, the deformation of a fluid body does not rotate in the same way as a rigid, aspherical deformation. In the case of a rigid body we would need to rotate for the full timestep $\Delta t$ around the spin vector axis given by $\vw$. In the fluid body case, deformations damp and fade away on the time scale $\tau$. Therefore, when $\Delta t/\tau \ll 1$, the body rotation has to be over a time step $\Delta t$, but when $\Delta t/\tau \gg 1$, then the rotation has to be over a time scale given by $\tau$. This constraint is secured by introducing the following time step for the rotational part:

\begin{equation}
\Delta t_{\rm{rot}} = \tau \left( 1 - \ep^{ -{\Delta t}/{\tau}} \right).
\label{eq:dtrot}
\end{equation}  

\noindent Given the spin vector, $\vw$,  and the rotational time step, $\Delta t_{\rm{rot}}$, we calculate the associated quaternion as

\begin{equation}
\vv{q} = (q_0,q_1,q_2,q_2) = \left[ \cos \left( \frac{\w \Delta t_{\rm{rot}}}{2} \right),  \sin \left( \frac{\w \Delta t_{\rm{rot}}}{2} \right) \frac{\vw}{\w} \right].
\label{eq:q}
\end{equation}

\noindent This quaternion is subsequently used to perform a rotation of the inertia tensor, by first calculating the rotation matrix
\be
\SR \left( {\bf q} \right) = 
\left[\begin{array}{ccc} 
2 \left( q_0^2 + q_1^2 \right) - 1   &  2 \left( q_1 q_2 - q_0 q_3 \right)   & 2 \left( q_1 q_3 + q_0 q_2 \right) \crm
2 \left( q_1 q_2 + q_0 q_3 \right)   &  2 \left( q_0^2 + q_2^2 \right) - 1   & 2 \left( q_2 q_3 - q_0 q_1 \right) \crm
2 \left( q_1 q_3 - q_0 q_2 \right)   &  2 \left( q_2 q_3 + q_0 q_1 \right)   & 2 \left( q_0^2 + q_3^2 \right) - 1
\end{array}\right] \ ,
\label{rot_matrix}
\ee
\noindent and then performing the actual rotation according to Eq.\,(\ref{160127a}). 
Note that we are not integrating quaternions, as for the deformation of fluid bodies it is unnecessary to keep track of the precise orientation of the body.

\subsection{Integration maps}\label{sec:integrators}

The conservation of angular momentum plays a key role in simulations of multi-body systems with tides. Tidal effects are usually small and therefore the angular momentum exchange between orbits and spins has to be calculated very accurately. The numerical error in the total angular momentum has to be kept to a minimum for the duration of the simulation. The time-symmetric, second-order Verlet-Leapfrog integrator exhibits excellent angular momentum conservation by design. The relative error in the angular momentum is kept close to machine-precision, and due to the symplectic nature of the algorithm, there is no drift in time. The base integrator of \texttt{TIDYMESS} is therefore the time-symmetric, second-order Verlet-Leapfrog (Drift-Kick-Drift) integrator:

\begin{empheq}[box=\widefbox]{align}
&  \vv{r}_{\frac{1}{2}} = \vv{r}_0 + \vv{v}_0 \frac{h}{2} & \rm{(Drift)} \nonumber \\
&  \vv{a}_{\frac{1}{2}}\left( \vv{r}_{\frac{1}{2}} \right) & \rm{(Acceleration)} \\
&  \vv{v}_1 = \vv{v}_0 + \vv{a}_{\frac{1}{2}} h & \rm{(Kick\,orbit)} \nonumber  \\
&  \vv{r}_1 = \vv{r}_{\frac{1}{2}} + \vv{v}_1 \frac{h}{2} & \rm{(Drift)} \nonumber
\end{empheq}

\noindent A fourth-order integrator is constructed by a composition of Leapfrog steps \citep[e.g.][]{Suzuki_1990, Yoshida_1993, McLachlan_1995}. We adopt a method by \citet{McLachlan_1995}, where a single ``McLachlan step'' of size $h$, is composed of 5 Leapfrog steps with sizes: $c_1 h$, $c_2 h$, $c_3 h$, $c_2 h$ and $c_1 h$, with $c_1 = 0.28$, $c_2 = 0.62546642846767004501$ and $c_3 = 1-2c_1-2c_2$. In combination with constant time steps, the pure $N$-body model in \texttt{TIDYMESS} is thus fourth-order and symplectic. 

For the conservative tidal model, we need to extend the Leapfrog step to include an update of the spin angular momenta using the torque. Furthermore, the accelerations and torques are not only functions of positions, but also of the inertia tensors: 

\begin{empheq}[box=\widefbox]{align}
&  \vv{r}_{\frac{1}{2}} = \vv{r}_0 + \vv{v}_0 \frac{h}{2} & \rm{(Drift)} \nonumber \\
&  \vv{a}_{\frac{1}{2}}\left( \vv{r}_{\frac{1}{2}}, \TI_{\frac{1}{2}} \right)  & \rm{(Acceleration)} \nonumber \\
&  \vT_{\frac{1}{2}}\left( \vv{r}_{\frac{1}{2}}, \TI_{\frac{1}{2}} \right)  & \rm{(Torque)} \\
&  \vv{v}_1 = \vv{v}_0 + \vv{a}_{\frac{1}{2}} h  & \rm{(Kick\,orbit)} \nonumber \\
&  \vL_1 = \vL_0 + \vT_{\frac{1}{2}} h  & \rm{(Kick\,spin)} \nonumber \\
&  \vv{r}_1 = \vv{r}_{\frac{1}{2}} + \vv{v}_1 \frac{h}{2}  & \rm{(Drift)} \nonumber
\end{empheq}

\noindent The problem here is that in the calculation of the acceleration and torque, the inertia tensor, $\TI_{\frac{1}{2}}$, is not known. In the conservative model, the inertia tensor can be split into a tidal part and a centrifugal part: $\TI_{\frac{1}{2}} = \TI_\eq\left(\vv{r}_{\frac{1}{2}}\right) + \TI_\eq\left( \vw_{\frac{1}{2}} \right)$. The positions, $\vv{r}_{\frac{1}{2}}$, are known, but not the spin frequencies, $\vw_{\frac{1}{2}}$. But in order to estimate $\vw_{\frac{1}{2}}$, through the relation $\vw_{\frac{1}{2}} = \TI^{-1}_{\frac{1}{2}} \vL_{\frac{1}{2}}$, we require $\TI_{\frac{1}{2}}$ again, thus producing an endless loop. Furthermore, $\vL_{\frac{1}{2}}$ is not known either. Inspired by the Auxiliary-Vector-Algorithm \citep[][AVA]{HellstromMikkola2009}, we resolve these issues by defining an auxiliary angular momentum, $\vK$. A reversible map is constructed as follows:

\begin{empheq}[box=\widefbox]{align}
&  \vv{r}_{\frac{1}{2}} = \vv{r}_0 + \vv{v}_0 \frac{h}{2} & \rm{(Drift)} \nonumber \\
&  \vT_{\frac{1}{2},0}\left( \vv{r}_{\frac{1}{2}}, \vL_0 \right)  & \rm{(Torque)} \nonumber \\
&  \vK_{\frac{1}{2}} = \vK_0 + \vT_{\frac{1}{2},0} \frac{h}{2}  & \rm{(Kick\,spin)} \nonumber \\
&  \vv{a}_{\frac{1}{2}}\left( \vv{r}_{\frac{1}{2}}, \vK_{\frac{1}{2}} \right)  & \rm{(Acceleration)} \nonumber \\
&  \vT_{\frac{1}{2}}\left( \vv{r}_{\frac{1}{2}}, \vK_{\frac{1}{2}} \right)  & \rm{(Torque)} \\
&  \vv{v}_1 = \vv{v}_0 + \vv{a}_{\frac{1}{2}} h  & \rm{(Kick\,orbit)} \nonumber \\
&  \vL_1 = \vL_0 + \vT_{\frac{1}{2}} h  & \rm{(Kick\,spin)} \nonumber \\
&  \vT_{\frac{1}{2},1}\left( \vv{r}_{\frac{1}{2}}, \vL_1 \right)  & \rm{(Torque)} \nonumber \\
&  \vK_1 = \vK_\frac{1}{2} + \vT_{\frac{1}{2},1} \frac{h}{2}  & \rm{(Kick\,spin)} \nonumber \\
&  \vv{r}_1 = \vv{r}_{\frac{1}{2}} + \vv{v}_1 \frac{h}{2}  & \rm{(Drift)} \nonumber
\end{empheq}

\noindent Note that for the acceleration and torque steps, we used the angular momentum as the argument instead of the inertia tensor. The inertia tensor is solved for iteratively. For example, the inertia tensor corresponding to $\vT_{\frac{1}{2},0}\left( \vv{r}_{\frac{1}{2}}, \vL_0 \right)$ is calculated as follows. We initially set $\vw = \vw_0$, and then we iterate:

\begin{empheq}[box=\widefbox]{align}
\begin{cases}
&  \TI = \TI_\eq\left( \vr_{\frac{1}{2}} \right) + \TI_\eq\left( \vw \right) \\
&  \vw = \TI^{-1} \vL_0 
\end{cases}
\end{empheq}

\noindent In \texttt{TIDYMESS}, the number of iterations can be set by the user, and is a balance between the required level of reversibility, and computational expense. In principle, with a sufficient number of iterations, the conservative tide model is symplectic. After performing several tests however, we noticed that the angular momenta, $\vL$ and $\vK$, can sometimes rapidly diverge. We resolve this by resetting $\vK=\vL$ after each ``McLachlan'' step. Although this affects the level of reversibility, in practice we still obtain excellent results. The final inertia tensor, $\TI_1$, is only calculated at the snapshot output interval, using the same iteration scheme.

The integration scheme for the linear tide model is very similar to that of the conservative model. The update of the inertia tensor now also includes a rotational component due to the tidal time lag, as well as the derivative of the equilibrium tensor, $\dot{\TI}^e$. We approximate the derivatives by 

\begin{empheq}[box=\widefbox]{align}
&  \dot{\TI}^e_{\frac{1}{2},0} = \left(\TI_\eq\left( \vr_{\frac{1}{2}}, \vL_0 \right)-\TI_\eq\left( \vr_0, \vL_0 \right)\right)\frac{2}{h} \nonumber \\
&  \dot{\TI}^e_{\frac{1}{2}} = \left(\TI_\eq\left( \vr_{\frac{1}{2}}, \vK_{\frac{1}{2}} \right)-\TI_\eq\left( \vr_0, \vL_0 \right)\right)\frac{2}{h} \\
&  \dot{\TI}^e_1 = \left(\TI_\eq\left( \vr_1, \vL_1 \right)-\TI_\eq\left( \vr_{\frac{1}{2}}, \vK_{\frac{1}{2}} \right)\right)\frac{2}{h} \nonumber 
\end{empheq}

The Creep tidal model with direct integration has a different integration map. Before the first integration step, we calculate the accelerations, $\vv{a}$, torques, $\vT$, and deformation tensor, $\dot{\TI}$. We also introduce the auxiliary inertia tensor, $\TP$. We subsequently integrate the system as follows:
\begin{empheq}[box=\widefbox]{align}
&  \vv{r}_{\frac{1}{2}} = \vv{r}_0 + \vv{v}_0 \frac{h}{2} & \rm{(Drift)} \nonumber \\
&  \vK_{\frac{1}{2}} = \vK_0 + \vT_0 \frac{h}{2}  & \rm{(Kick\,spin)} \nonumber \\
&  \TP_{\frac{1}{2}} = \TP_0 + \dot{\TI}_0 \frac{h}{2}  & \rm{(Kick\,shape)} \nonumber \\
&  \vw_{\frac{1}{2}} = \TP_{\frac{1}{2}}^{-1} \vK_{\frac{1}{2}} & \rm{(Spin\,frequency)} \nonumber \\
& \noindent\rule{2cm}{0.4pt} & \nonumber \\
&  \vv{a}_{\frac{1}{2}}\left( \TP_{\frac{1}{2}}, \vv{r}_{\frac{1}{2}} \right)  & \rm{(Acceleration)} \nonumber \\
&  \vT_{\frac{1}{2}}\left( \TP_{\frac{1}{2}}, \vv{r}_{\frac{1}{2}} \right)  & \rm{(Torque)} \nonumber \\
&  \dot{\TI}_{\frac{1}{2}}\left( \TP_{\frac{1}{2}}, \vv{r}_{\frac{1}{2}}, \vw_{\frac{1}{2}} \right)  & \rm{(Deformation)} \nonumber \\
& \noindent\rule{2cm}{0.4pt} & \nonumber \\
&  \vv{v}_1 = \vv{v}_0 + \vv{a}_{\frac{1}{2}} h  & \rm{(Kick\,orbit)} \nonumber \\
&  \vL_1 = \vL_0 + \vT_{\frac{1}{2}} h  & \rm{(Kick\,spin)}  \\
&  \vv{\TI}_1 = \vv{\TI}_0 + \dot{\TI}_{\frac{1}{2}} h  & \rm{(Kick\,shape)} \nonumber \\
&  \vv{r}_1 = \vv{r}_{\frac{1}{2}} + \vv{v}_1 \frac{h}{2}  & \rm{(Drift)} \nonumber \\ 
& \vw_1 = \TI_1^{-1} \vL_1 & \rm{(Spin\,frequency)} \nonumber \\
& \noindent\rule{2cm}{0.4pt} & \nonumber \\
&  \vv{a}_1\left( \TI_1, \vv{r}_1 \right)  & \rm{(Acceleration)} \nonumber \\
&  \vT_1\left( \TI_1, \vv{r}_1 \right)  & \rm{(Torque)} \nonumber \\
&  \dot{\TI}_1\left( \TI_1, \vv{r}_1, \vw_1 \right)  & \rm{(Deformation)} \nonumber \\
& \noindent\rule{2cm}{0.4pt} & \nonumber \\
&  \vK_1 = \vK_{\frac{1}{2}} + \vT_1 \frac{h}{2}  & \rm{(Kick\,spin)} \nonumber \\
&  \TP_1 = \TP_{\frac{1}{2}} + \dot{\TI}_1 \frac{h}{2}  & \rm{(Kick\,shape)} \nonumber
\end{empheq}

\noindent A consequence of directly using the derivative of the inertia tensor, is that the time step cannot be made too large, since it has to resolve the orbital, spin and deformation time scales. This becomes problematic for rapidly spinning bodies, or tides in the linear regime. As a resolution to this limitation, we implement a tidal model based on the general and optimized approximation of the Creep model (see Sec.~\ref{sec:new_method}). The integration map is the same as for the linear tide model, both requiring the equilibrium tensor at the beginning and end of a time step. The net deformation is split into a tidal deformation and rotational part. In the positive time direction, we first perform the tidal deformation followed by the rotation. In the negative time direction, the inverse sequence is applied. 

Although the tidal deformation equation (Eq.\,(\ref{eq:Ifinal2})) is time reversible, in a numerical sense it is not. For values of $\Delta t / \tau \gg 1$, the exponentials become small enough, such that any memory of initial conditions gets lost in the numerical noise. This affects the performance of the fourth-order composition of the integrator, which includes a negative time step. In order to resolve this problem, we check the ratio of $\Delta t / \tau$ for each individual body. If this ratio is larger than some critical value, $f_{\rm{crit}}$, than we use the equation from the linear tide model instead. We find that a value of $f_{\rm{crit}} = 7$ to be optimal.    

\subsection{Time step functions}

\texttt{TIDYMESS} implements three time step criteria: 

\begin{itemize}
\item Constant
\item Adaptive,
\item Adaptive, weighted and symmetrised.
\end{itemize}

\noindent In the adaptive case, we adopt the minimum of the pairwise free fall and interparticle flyby time scales \citep[e.g.][]{Pelupessy_Janes_2012}. The magnitude of the time steps can be controlled by the constant time step multiplicative factor, $\eta$. 

Adapting the time step size affects the advantageous conservation properties of symplectic integrators. One method to improve this is by symmetrising the time steps \citep[e.g.][]{Dehnen_2017}. Symmetrisation methods are compatible with time step functions which are smooth and continuous. The latter can be accomplished by using adaptive time steps which are weighted over all pairs. We adopt the geometric symmetrisation method in combination with the direct averaging method, i.e. method B1g of \citet{TB_ACM_22}.   
For the direct Creep model, we extend the pairwise, orbital time steps to include spin periods and fluid relaxation time scales. The weighting method however, is only applied to the orbital time step functions. 

\subsection{Input/Output}

\texttt{TIDYMESS} reads in two files: the parameter file and the initial condition file. Their default names are \texttt{tidymess.par} and \texttt{tidymess.ic}, but these can be user defined. The parameter file specifies various input parameters necessary to run a simulation. Here, the user defines which tidal model to use (Table~\ref{tab:tidal_models}), the Post-Newtonian order (Sect.~\ref{section:grc}), whether to turn on magnetic braking effects (Sect.~\ref{section:mbe}), and how to handle collisions (Sect.~\ref{section:col}). The simulation time is specified here, as well as the time step function and time step parameter. The user can specify the output directory, the snapshot time interval, and the unit system to be used in the  output snapshots. Various output formats are available, including an output file per body, output file per snapshot, or all snapshots inside a single file. The snapshots are always in Cartesian coordinates in the inertial frame specified by the initial condition. 
Other functionalities include a flag for restarting a simulation, setting the initial shapes of the bodies (sphere or equilibrium shape), varying the number of bodies to include from the initial condition file, specifying the type of coordinates used in the initial condition file, for positions and velocities (Cartesian, elliptical astrocentric or elliptical Jacobian), as well as for spins (which can be given in the inertial frame or with respect to the bodies' orbital plane around the primary body with mass $\M$). The number of iterations to improve the reversibility of the integrator can also be defined (see Sec.~\ref{sec:integrators}). All input parameters can also be defined on the terminal command line. In case parameters are defined in both the parameter file and the terminal, then the latter value is assigned.   

The initial condition file consists of a table with the properties of the bodies in the system. The internal properties (mass $\m$, mean radius $R$, moment of inertia factor $\xix$, fluid Love number $\kf$, fluid relaxation time $\taub$, and magnetic braking constant $\alpha$) can all be set in various physical units. The positions and velocities, as well as the initial spin vectors,  can be defined here, in the coordinate frame specified in the parameter file (Cartesian coordinates or orbital elements). Each body can also be assigned a name tag. Hence, in order to start a new simulation, the user must first gather the necessary data from an external source, and convert that to a table of initial conditions in the \texttt{TIDYMESS} format. Then, the user updates the parameter file in order to define the simulation that they wish to run. The outputted snapshots are text files, which are easily read by other programs, such as \texttt{PYTHON}. Binary files are also written out for backup and restart purposes. A summary log file is also kept, with information such as the duration of the simulation, and the level of conservation of various quantities.   




\section{Demonstrations of \texttt{TIDYMESS}}\label{sec:results}

In this section we provide six demonstrations of \texttt{TIDYMESS}, each highlighting a different tidal feature or application. In Sec.~\ref{sec:ex1}, we consider the current-day Earth-Moon system, and model the orbital recession speed of the Moon due to tides. In Sec.~\ref{sec:ex2}, we focus on the deformations of the heartbeat star KOI-54. In Sec.~\ref{sec:ex3}, we model the eccentric exoplanet HD80606b, and compare to previous results from the literature. In Sec.~\ref{sec:ex4}, we consider the three-body problem consisting of the early Sun-Earth-Moon system, and we reproduce the evection and eviction events. In Sec.~\ref{sec:ex5}, we model a resonant chain of three planets, and the formation of a hot Jupiter. In Sec.~\ref{sec:ex6}, we consider the tidal evolution of a stellar association, until the formation of the first, synchronised binary system. We note that the aim of these examples is to provide a gallery of demonstrations of \texttt{TIDYMESS}, rather than providing in-depth analyses of the results, which can mostly be found in the given references.

\subsection{Earth-Moon recession speed}\label{sec:ex1}

In our first demonstration of \texttt{TIDYMESS}, we consider the current-day Earth-Moon system. We simplify the system to a one-body system, i.e. we treat the Moon as a point-mass body and only consider tides on Earth. We simplify the problem somewhat more by assuming a zero obliquity, and a circular orbit. The initial conditions are taken from NASA's fact sheets \citep{nasa} and given in Tab.~\ref{tab:earth_moon}. We evolve the system for 1000 years, and measure the semi-major axis as a function of time. The slope of a linear fit gives the average recession speed of the Moon from the Earth. \texttt{TIDYMESS} also outputs a log file, including the simulation wall-clock time. 

\begin{figure}
\centering
\begin{tabular}{c}
\hspace{-0.5cm}
\includegraphics[height=0.63\textwidth,width=0.45\textwidth]{\figpath 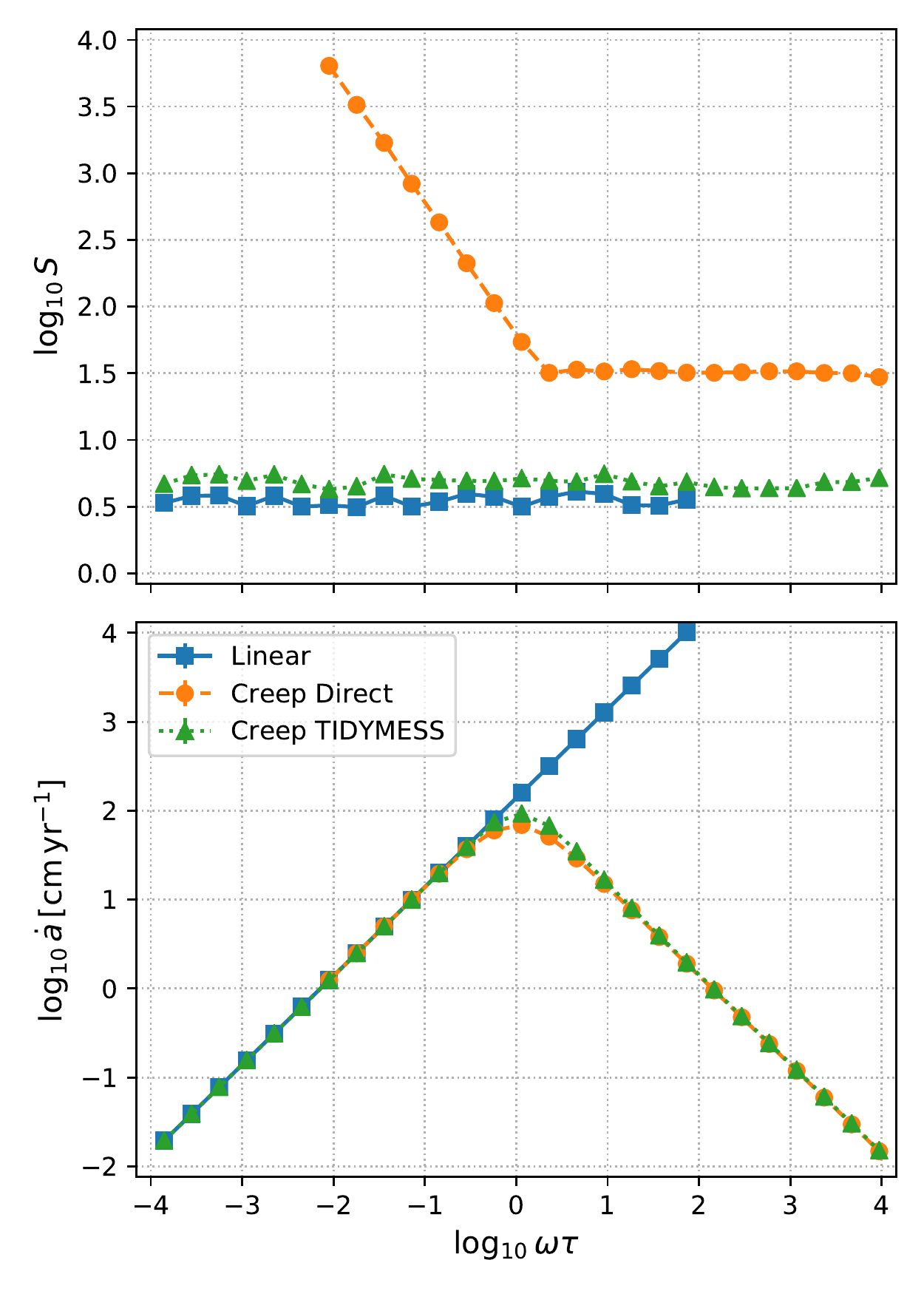} \\
\end{tabular}  
\caption{ Tides in the current Earth-Moon system. As a function of the tidal parameter, $\delta = \omt \tau$, we plot the slow down factor of the wall-clock time, relative to simulations without tides (top panel), and the recession speed of the Moon in terms of the derivative of the semi-major axis, $\dot{a}$ (bottom panel). For the Creep models, we observe a maximum recession speed at $\omt \tau = 1$. We confirm the linear model is consistent for $\omt \tau \ll 1$, but blows up in the opposite regime. This example also illustrates the performance of the \texttt{TIDYMESS} creep model, which is independent of the fluid relaxation time. This is in contrast to the direct creep model, which becomes excessively expensive in the linear regime. }
\label{fig:earth_moon}
\end{figure}

In Fig.~\ref{fig:earth_moon}, we plot the recession speed as a function of the tidal parameter, $\delta = \omt \tau$. Here, $\omt$ is the tidal forcing frequency given by $\omt = 2\left( \w - n \right)$, with $n$ the orbital frequency (mean motion). Starting with the linear tide model, we observe a linear increase of the recession speed of the Moon as a function of tidal parameter. Intuitively, a larger tidal parameter corresponds to a larger lag of the tidal bulge, resulting in a larger torque. In the linear model, the increase in the recession speed is maintained, even up to very large tidal parameters. However, in the limit of infinite fluid relaxation times, the shapes of the bodies do not change, thus resulting in a conservative system without tidal dissipation. Hence, there should be a non-linear regime in which the tidal dissipation (and therefore the recession speed) decreases again. The Creep models exhibit such a feature. The peak recession speed occurs when the forcing frequency resonates with the fluid relaxation time, i.e. $\omt \tau = 1$. The curve is symmetric around this peak. We confirm that in the linear regime, the Creep model reduces to the linear tide model. Furthermore, the two Creep model implementations are mutually consistent, except for a small deviation near the peak of the curve. This follows from the approximations made in the Creep \texttt{TIDYMESS} algorithm. We also ran the simulations with the conservative tidal model, and we confirm a negligible residual recession speed $\leq 10^{-5}\,\rm{[cm\,yr^{-1}]}$, independent of the tidal parameter. 

A large difference is observed in the performance of the two Creep models. In Fig.~\ref{fig:earth_moon}, we also plot the wall-clock time of the simulation, normalised by the wall-clock time for the pure $N$-body model. The linear model is about a factor 3 slower than a normal $N$-body code. The Creep \texttt{TIDYMESS} model is 4-5 times slower\footnote{This benchmark was obtained using 1 iteration in the integration map. Without iterations, the \texttt{TIDYMESS} Creep model is 3-4 times slower.}, but the performance is independent of the tidal parameter. The direct Creep model is about 30 times slower in the non-linear regime, which is explained by the rapid spin of Earth compared to its orbit (1 month / 1 day $\approx 30$). However, into the linear regime, the wall-clock time increases linearly with decreasing fluid relaxation time scale. This example demonstrates the advantage of the new \texttt{TIDYMESS} implementation of the Creep model.  

\begin{figure}
\centering
\begin{tabular}{c}
\includegraphics[height=0.72\textwidth,width=0.45\textwidth]{\figpath 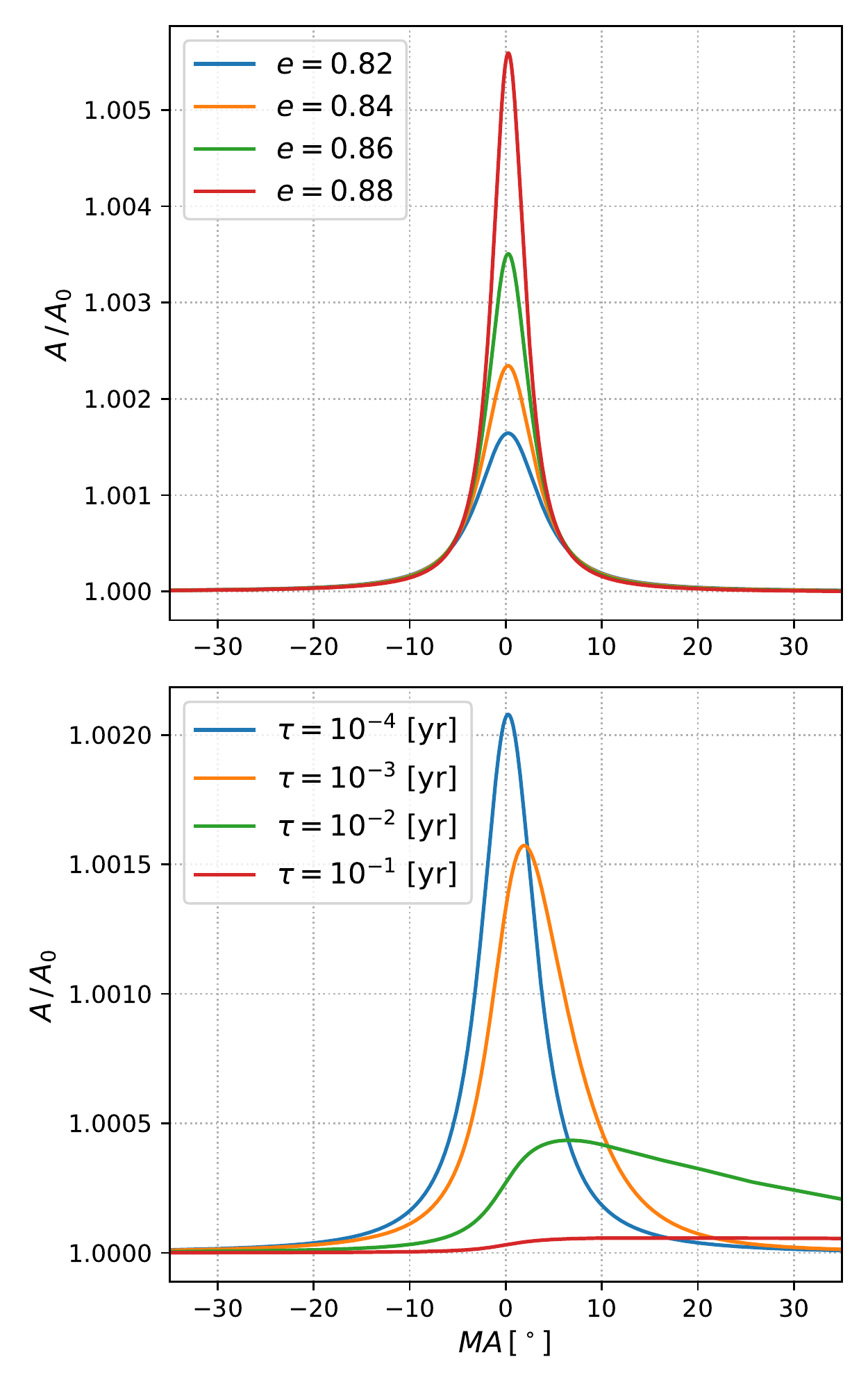} \\
\end{tabular}  
\caption{Heartbeat of the eccentric binary system KOI-54. We plot the normalised, projected surface area of the stars during pericenter passage. The normalisation is taken to be the initial value at apocenter starting from their equilibrium shapes, which is not necessarily the same as the sum of the two unperturbed stars. The profile depends sensitively on the eccentricity (top panel, $\tau = 10^{-4}\,\rm{[yr]}$), as well as the fluid relaxation time (bottom panel, $e=0.8335$). The latter introduces an asymmetry in the profile.  }
\label{fig:heartbeat}
\end{figure}

\subsection{KOI-54: eccentric binary with a heartbeat}\label{sec:ex2}  
  
\begin{figure*}
\centering
\begin{tabular}{c}
\includegraphics[height=1.045\textwidth,width=0.95\textwidth]{\figpath 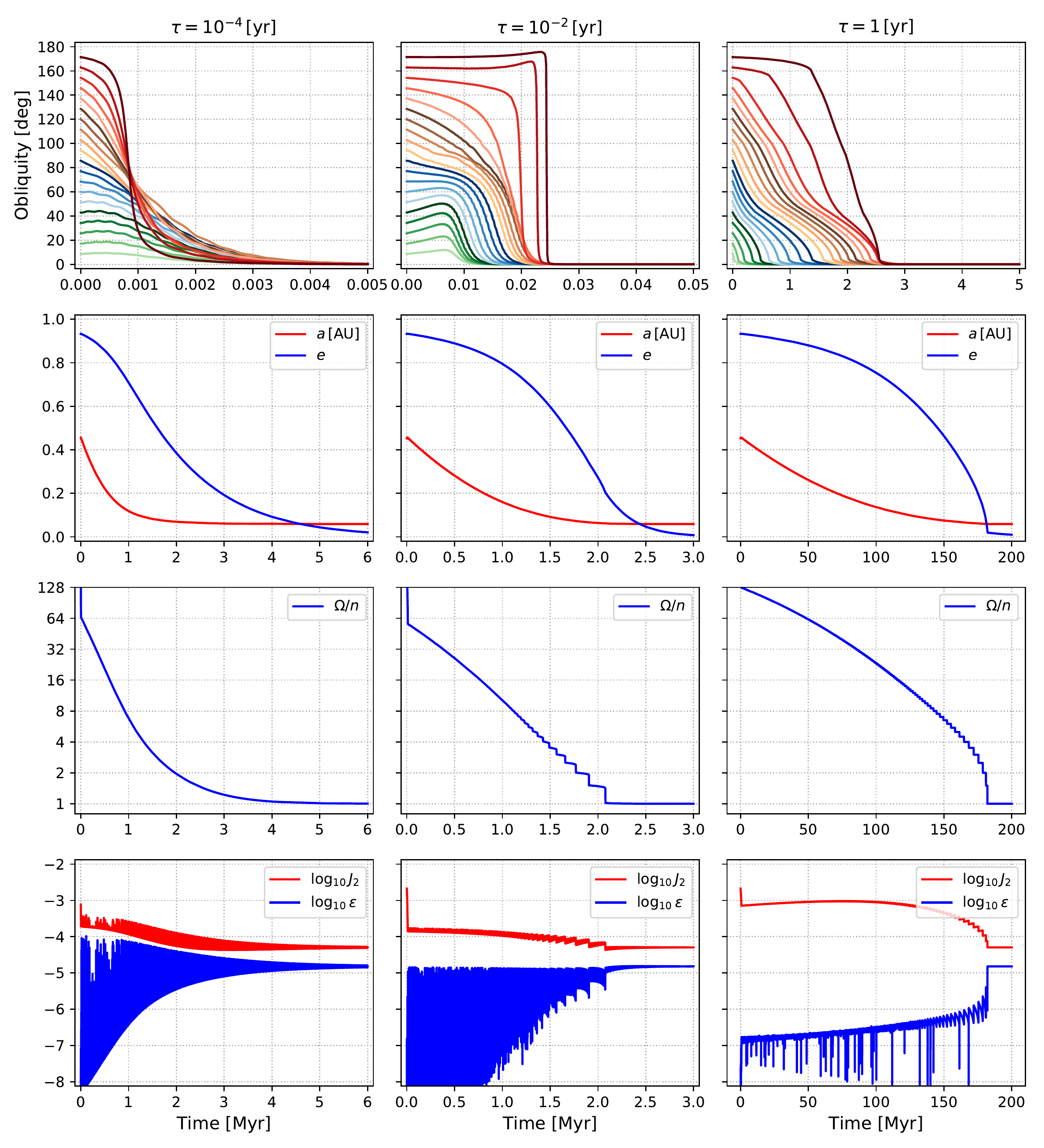} \\
\end{tabular}  
\caption{ Tidal evolution of exoplanet HD80606b. We plot the time evolution of various initial obliquities (top panel), semimajor-axis and eccentricity ($2^{\rm{nd}}$ row), ratio of spin to orbital frequency ($3^{\rm{rd}}$ row), and two parameters describing oblateness ($J_2$) and prolateness ($\epsilon$) of the planet (bottom panel). We vary the fluid relaxation time per column, such that we sample different tidal regimes: linear (left), peak dissipation (middle) and non-linear/inverse (right). In each case, the planet attains the synchronised, equilibrium state.  }
\label{fig:hd80606b}
\end{figure*}
  
Apart from the usual phase space coordinates, \texttt{TIDYMESS} also integrates and outputs the spin and inertia tensor coordinates in an inertial reference frame. This makes it possible to project the extended bodies onto a plane, and measure their projected surface areas as a function of time. 

Heartbeat stars are known for their periodic flux increases. They consist of eccentric binary stars, and during pericenter passage, tidal deformations cause a brief spike in the observed flux. One observed example is the system KOI-54 \citep[e.g.][]{Welsh_2011}. We adopt their derived system parameters, which are repeated in Tab.~\ref{tab:koi54}. We perform brief simulations for a single orbital period, using the direct Creep tidal model. The projected surface area is calculated according to the method described in appendix~\ref{app:A}.  

In Fig.~\ref{fig:heartbeat}, we plot the normalised increase of the projected surface area as a function of mean anomaly. We first fix the fluid relaxation time of the stars to $\tau = 10^{-4}\,\rm{[yr]}$, and vary the eccentricity (top panel). We confirm the finding by \citet{Welsh_2011}, that the flux (which is proportional to surface area), depends sensitively on the eccentricity. This is explained by the steep dependence of tides on separation. Secondly, we fix the eccentricity to $e=0.8335$, and systematically vary the fluid relaxation time (bottom panel). We observe that the fluid relaxation time of the stars affects the symmetry of the profile. The larger the value of $\tau$, the longer it takes for the star to restore to its non-perturbed shape after pericenter passage. 
Detailed observations of stellar shapes thus hold stringent information on their tidal response. 
 
\subsection{HD80606b: a very eccentric Jupiter}\label{sec:ex3}

The exoplanet HD80606b is highly eccentric and  its future tidal evolution will most likely evolve it into a hot Jupiter. Here, we reproduce  simulations by \citet{Correia_etal_2014} and \citet{Boue_etal_2016}, who adopted the same tidal creep model as \texttt{TIDYMESS}. The initial conditions are repeated in  Tab.~\ref{tab:hd80606b}. 
We evolve the planet HD80606b with three different values for the fluid relaxation time: $\tau = 10^{-4}, 10^{-2}$ and 1 year. These values represent the linear regime, the non-linear regime, and the transitional regime in between.  

Starting with a zero obliquity, we evolve the systems until tidal synchronisation is approximately reached. We confirm in Fig.~\ref{fig:hd80606b} ($2^{\rm{nd}}$ row), that the semi-major axes shrink until a fixed value determined by angular momentum conservation. The eccentricities also decrease to zero. At the same time, we observe that the ratio between the spin and orbital frequencies evolves to unity (Fig.~\ref{fig:hd80606b}, $3^{\rm{rd}}$ row). We successfully reproduce the ``staircase'' pattern found by \citet{Correia_etal_2014}, which is a consequence of passing through a spectrum of spin-orbit resonances in the non-linear regime. The deformation of the planet evolves over a similar time scale as the orbits and spins. We measure the planet's Stokes coefficients $J_2$ and $\epsilon = (C_{22}^2 + S_{22}^2)^{1/2}$, which represent the oblateness and prolateness, respectively (Fig.~\ref{fig:hd80606b}, bottom row). Large variations are observed in the prolateness due to the eccentric orbit and the steep dependence of tides on separation. Towards the end, an equilibrium figure is reached, which is approximately independent of fluid relaxation time. A non-zero initial obliquity rapidly decreases to zero on a time scale much shorter than the orbital time scale (Fig.~\ref{fig:hd80606b}, top row). Hence, large initial obliquities do not change the subsequent evolution much. 

\subsection{Early evolution of the Sun-Earth-Moon system}\label{sec:ex4}

The Moon is currently receding from Earth with a rate of about 4~$\rm{cm\,yr^{-1}}$ \citep{Dickey_etal_1994}. This implies that in the past the Moon was closer to Earth. Origin scenarios, such as a giant impact \citep[e.g][]{Canup_2004}, indeed consider an ``initial'' orbit after formation, just beyond the Roche limit. Supersynchronous rotation of the Earth and Moon subsequently resulted in an orbital expansion. During this expansion, multiple interesting resonances are crossed, which excite the eccentricity and the inclination.    

We reproduce the results from \citet{Touma_1998}, who analyze the resonant phenomena of evection and eviction. In the case of evection, the eccentricity of the Earth-Moon orbit is excited due to a resonance between Earth's orbital frequency around the Sun and the Moon's pericenter precession rate. This is followed by eviction, where the inclination of the Moon is excited due to an eccentricity-inclination resonance \citep{Touma_1998}. These two resonance passages provide a stringent validation test for \texttt{TIDYMESS}. 

We adopt the initial conditions from \citep{Touma_1998}, which are repeated here in  Tab.~\ref{tab:sun_earth_moon}. The Moon is initially in the equatorial plane of the Earth in a near circular orbit. The Earth-Moon separation starts off at 3.5 Earth radii, while Earth's obliquity with respect to its orbit around the Sun is set to 10 degrees. The initial rotation period of Earth is chosen to be 5.0~hr. In our experiment, we will only consider the tidal deformation of the Earth, whose fluid relaxation time scale is set to one second, i.e. linear tidal regime. 

\begin{figure}
\centering
\begin{tabular}{c}
\includegraphics[height=0.72\textwidth,width=0.45\textwidth]{\figpath 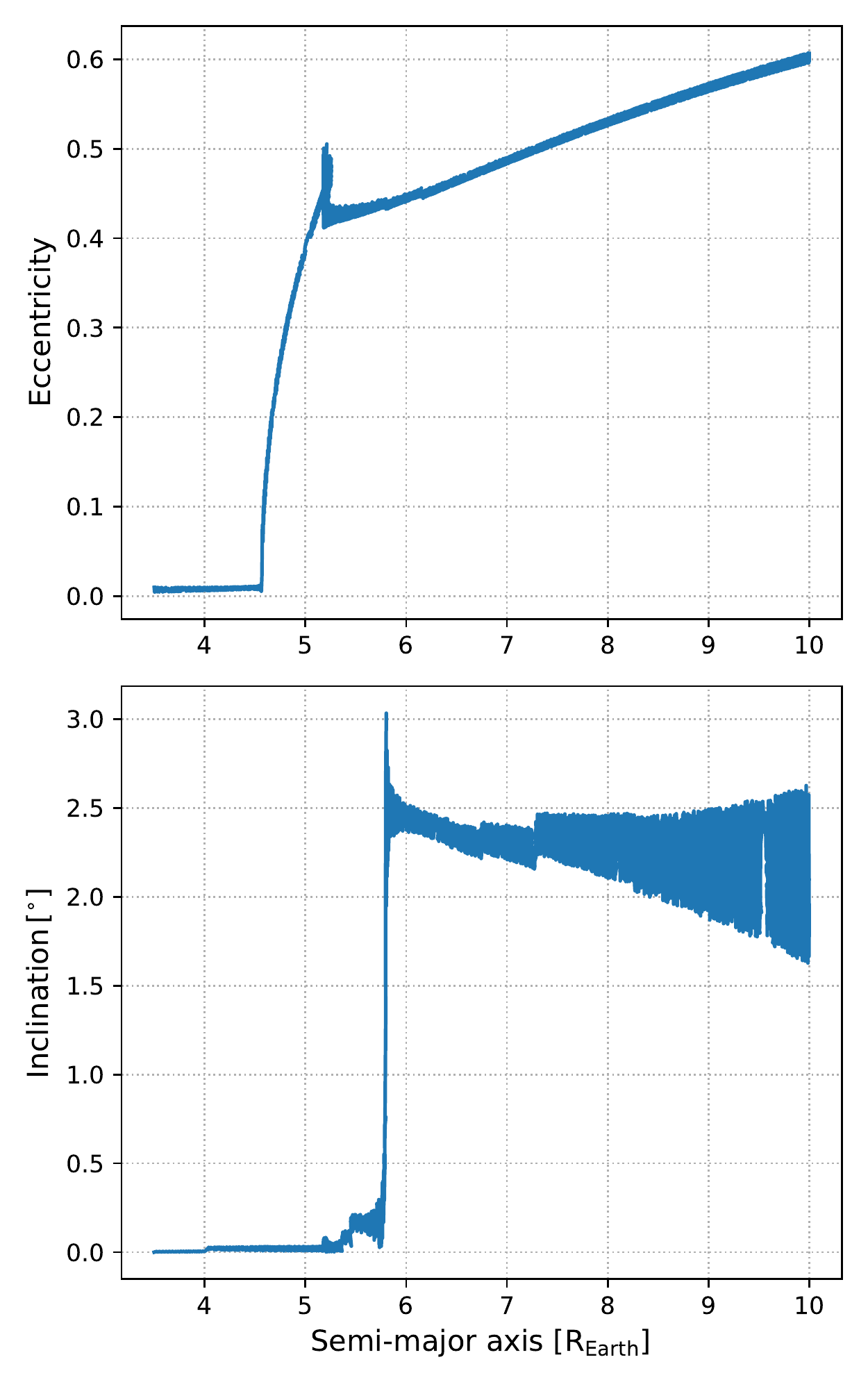} \\
\end{tabular}  
\caption{Evection resonance leads to eccentricity excitation (top), and eviction resonance leads to inclination excitation (bottom), during the early evolution of the Sun-Earth-Moon system. These results are consistent with those from \citet{Touma_1998}.  }
\label{fig:evection}
\end{figure}

In Fig.~\ref{fig:evection}, we plot the evolution of the eccentricity and inclination as a function of semi-major axis for the Earth-Moon orbit. We reproduce the increase of the eccentricity to a value of about 0.5, near a separation of $\sim 4.5$ Earth radii. We also confirm that the eviction phase occurs at a separation between 5.5 and 6 Earth radii, resulting in an increase in the inclination of the Moon to a value between 2 and 3 degrees. Hence, \texttt{TIDYMESS} can be used to model resonant phenomena in the Solar System.

\subsection{Chaotic planet-planet scattering}\label{sec:ex5}  

\begin{figure*}
\centering
\begin{tabular}{c}
\includegraphics[height=0.7\textwidth,width=0.9\textwidth]{\figpath 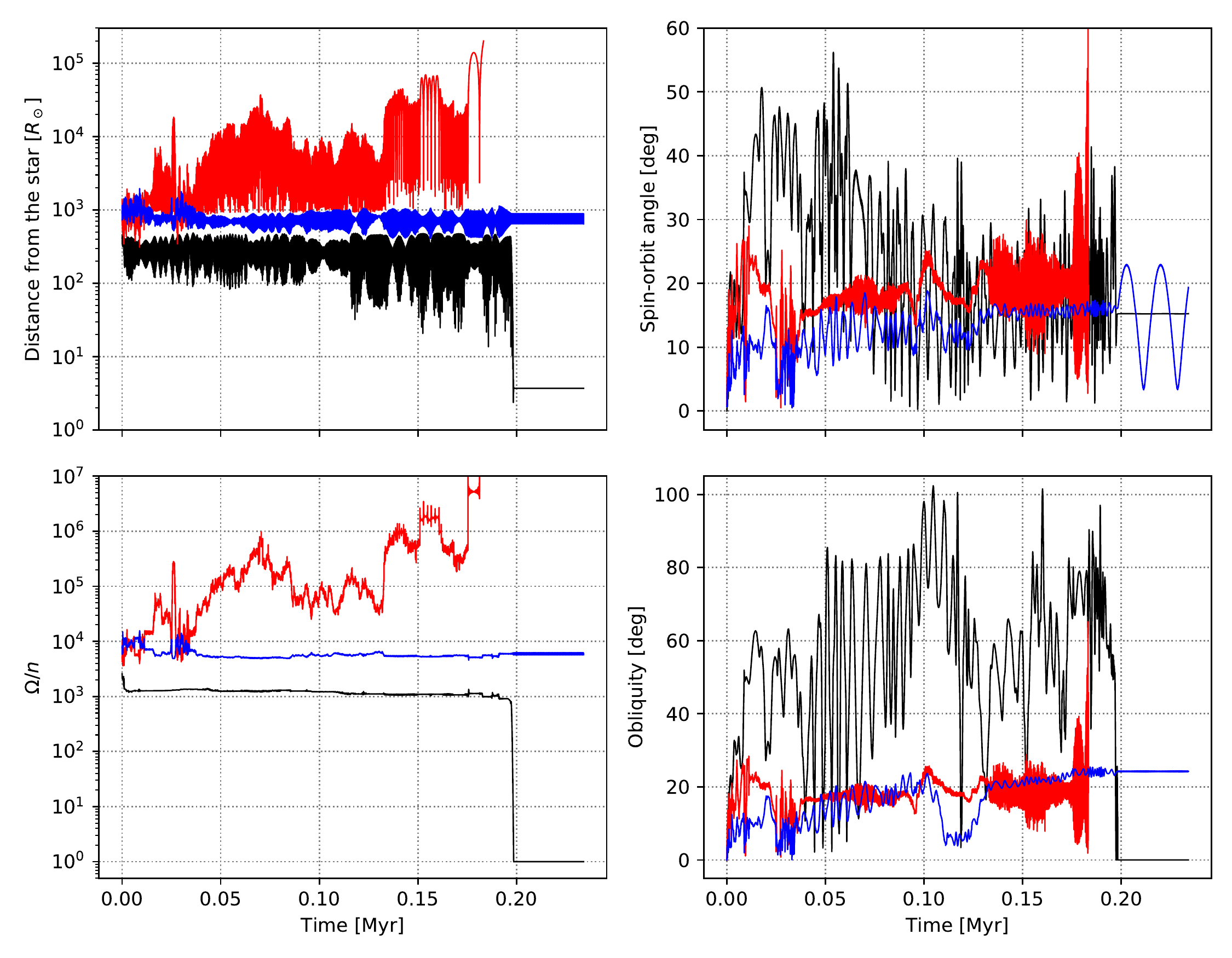} \\
\end{tabular}  
\caption{We plot the chaotic orbital separations of three planets from their host star (top left). One planet is ejected after about 0.18 Myr (red curve), while another planet is scattered inwards, after which its orbit is tidally synchronised, i.e. it has become a hot Jupiter (black curve). This planet's final spin-orbit angle is about 16 degrees (top right), its spin frequency is synchronised to its orbital frequency (bottom left), and its obliquity is zero (bottom right). The planet with the relatively unaffected orbital separation (blue curve), obtains a final obliquity of about 24 degrees. This demonstrates that planet-planet scattering can change the obliquities of planets. This planet also exhibits an oscillation of the spin-orbit angle (top right, blue curve), which is explained by the spin precession and nutation of the host star, due to its tidal coupling with the hot Jupiter.   }
\label{fig:planet_scatter}
\end{figure*}

\texttt{TIDYMESS} covers a niche regime of parameter space where orbits are chaotic and tides can become non-linear. As a demonstration, we consider a host star and a three-planet system, which are all mutually tidally interacting. If the planets start out in a resonant chain, i.e. each neighbouring pair of planets is in orbital resonance, instabilities can grow and excite eccentricities. When two planets are nearly orbit-crossing, strong gravitational deflections will lead to a variety of possible outcomes. These include planet ejection, rearrangement of orbits, giant impacts, and the formation of a hot Jupiter \citep[e.g.][]{Beauge_Nesvorny_2012}. In the latter case, one planet is scattered towards the host star. If the perihelion distance is of order a few times the stellar radius, then the planet can be tidally captured by the star. The subsequent evolution might resemble that of exoplanet HD80606b (see  Sec.~\ref{sec:ex3}). We note that \texttt{TIDYMESS} can model both the initial multi-planet interaction phase, as well as the subsequent tidal synchronisation of the hot Jupiter, without any need for a transition between numerical methods. Furthermore, it is possible to model bodies in the same system with different tidal models. In our experiment, we will assign the host star the conservative tide model, while the planets respond according to the \texttt{TIDYMESS} Creep model. 

We define our own initial condition for this experiment, which might be common among young exoplanetary systems. The host star is taken to be a Sun-like star. The planets have masses 1, 2 and 4 times the mass of Jupiter, with the lightest planet nearest the star, and the heaviest planet on the outside. The resonant chain consists of a 2:1 and a 3:2 orbital resonance between the inner and outer pair respectively. The inner planet has a semi-major axis of 2\,AU. 
The spin periods of the bodies are taken to be the current-day values of the Sun and Jupiter.
Due to the presence of chaos, small changes in the initial realization have a big influence on the final outcome. We randomly varied the eccentricities between 0 and 0.04, and inclinations between $\pm 0.4$ degrees. The initial orbital phases were also randomly sampled. For our demonstration, we chose a system which produced both a hot Jupiter and a planet ejection. 

In Fig.~\ref{fig:planet_scatter}, we plot the time evolution of various spin-orbit properties of the three planets. We observe in the top left panel that the distance of the planets to the host star varies chaotically until about 0.18\,Myr, when one planet is ejected from the system (red curve, initially the middle planet). Some time later just before 0.20\,Myr, a planet is scattered inwards, followed by a rapid tidal synchronisation with the host star (black curve, initially the inner and lightest planet). The final semi-major axis of the hot Jupiter is about 4-5 stellar radii. The initial outer planet, which is also the highest mass, remains relatively unaffected in its orbit. 

In parallel to the orbital evolution, the hot Jupiter also synchronises its spin (bottom left panel), while the spin-orbit angle becomes fixed at approximately 16 degrees (top right panel). The hot Jupiter's obliquity was greatly excited during the multi-planet scattering phase, up to about 100 degrees. Tidal evolution after the scatter however, synchronises the obliquity rapidly to zero degrees (bottom right panel). Interestingly, the most massive planet obtains a final obliquity of about 24 degrees (blue curve, bottom right panel). This result shows that planet-planet scattering is a channel for producing planets with large obliquities. Furthermore, the same planet shows large variations in the final spin-orbit angle by about 20 degrees. This is explained by the spin precession and nutation of the star, which in turn is driven by the tidal coupling between the star and the hot Jupiter.

\subsection{Close binary formation in a stellar association}\label{sec:ex6}  

\begin{figure}
\centering
\begin{tabular}{c}
\includegraphics[height=0.72\textwidth,width=0.45\textwidth]{\figpath 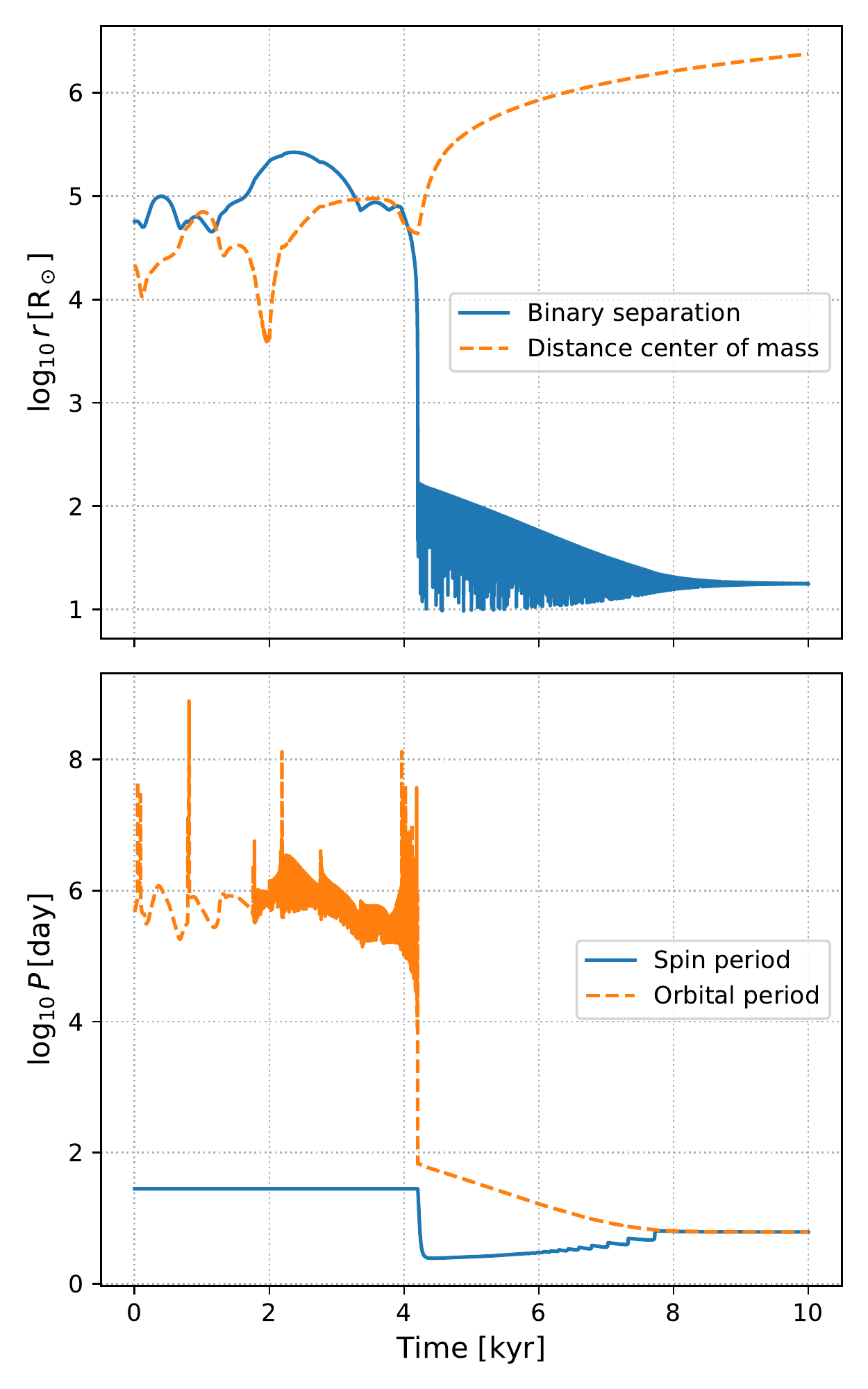} \\
\end{tabular}  
\caption{Formation of a close binary star from  a stellar association. A three-body encounter produces an ejected binary star after about 4\,kyr. This eccentric binary subsequently tidally evolves into a circular and synchronised binary star with an orbital period of about 6 days. }
\label{fig:plummer}
\end{figure}

\texttt{TIDYMESS} can model the tidal evolution of an arbitrary configuration of stars and planets. These include moons around planets, a host star with multiple planets, binary stars with planets, hierarchical triple stars, or even stellar associations up to a few hundred stars.\footnote{A future version of \texttt{TIDYMESS} is planned to have a parallel implementation allowing larger values for $N$. }
Tidal effects play an important role in stellar systems, and can result in the formation of binary stars by tidal capture \citep[e.g.][]{Press_1977}, and potentially rearrange stellar spins.

Here, we will consider a stellar association consisting of 32 single, solar mass stars in a Plummer configuration \citep{Plummer_1911}. This is the minimal number of stars for a system with two-body relaxation effects \citep{Zwart_2021}. 
The virial radius of the association is set to $10^5$ solar radii (for denser systems, we observed the common occurrence of collisions, i.e. formation of blue stragglers).  
The tidal model is the \texttt{TIDYMESS} Creep model, with a fluid relaxation time arbitrarily set to 0.01 year (3.65 days), which is of the same order as the orbital period of very close binary stars. We evolve the association for 10\,kyr (approximately 12 dynamical times). 

In Fig.~\ref{fig:plummer}, we visualise the formation of a close binary star through a two-part process. Up to about 4\,kyr, the stars are moving chaotically through the association, ocassionally encountering other stars. The separation between the two stars which eventually form the binary, varies chaotically around a value given by the size of the association ($10^5$ solar radii, blue curve in top panel). Then a three-body encounter occurs, resulting in the ejection of a star and an eccentric binary. In Fig.~\ref{fig:plummer}, we observe that the distance of the binary from the center of mass of the association starts to linearly increase after 4\,kyr (top panel, orange curve). At the same time, the tidal evolution of the binary results in orbital shrinkage and circularisation. The final orbit has an orbital period of about 6 days. For the two binary components, we also measure the spin periods and orbital periods in time (bottom panel). The spins of both stars are indistinguishable. We estimate the orbital period in the initial unbound phase as the separation divided by the relative speed times $2\pi$. In the initial phase up to 4\,kyr, the stars are supersynchronous. Spins are greatly affected however in the binary system, through tidal synchronisation. The spins of the remaining stars in the association remained barely affected on this short time scale. 
It remains an open question whether tides in large-$N$ stellar systems can redistribute the spin vector distribution.


\section{Conclusion}\label{sec:conclusions}

We present the new $N$-body code \texttt{TIDYMESS}, which stands for the ``TIdal DYnamics of Multi-body ExtraSolar Systems''. Due to the combination of a direct integration method and the tidal Creep model, this code covers a niche regime of parameter space where orbits become chaotic and where tides can be non-linear. 
A new implementation of the Creep model, which we call the \texttt{TIDYMESS} Creep model, consists of an integration method for the tidal deformations, which is independent of the spin and fluid relaxation time scales. This greatly benefits the performance of the Creep model in the linear regime, or in the presence of rapidly spinning bodies. 
 
We provide six different demonstrations of \texttt{TIDYMESS}, with applications to the tidal evolution of moons, planets and stars. The code will be made open-source and provide the community with a new method for studying the chaotic, tidal evolution of Solar System bodies, exoplanets,  stellar multiples and associations. Comparison to other tidal $N$-body codes, combined with stringent constraints from current and upcoming observations, will allow us to scrutinise the physics of tidal dissipation and its manifestations in various astronomical bodies. 


\subsection*{Acknowledgements}

This work was supported by
funds from the European Research Council (ERC) under the European Union’s Horizon 2020 research and innovation program under grant agreement No 638435 (GalNUC). Other support includes
CFisUC (UIDB/04564/2020 and UIDP/04564/2020),
GRAVITY (PTDC/FIS-AST/7002/2020),
PHOBOS (POCI-01-0145-FEDER-029932), and
ENGAGE~SKA (POCI-01-0145-FEDER-022217),
funded by COMPETE 2020 and FCT, Portugal.
The numerical simulations were performed 
at the Laboratory for Advanced Computing at University of Coimbra (\href{https://www.uc.pt/lca}{https://www.uc.pt/lca}).
The diagrams were produced using \texttt{PYTHON} \citep{van1995python}. 

\section*{Data Availability}

The data underlying this article will be shared on reasonable request to the corresponding author.

\bibliographystyle{mn2e}
\bibliography{tidymess-S}   


\appendix

\section{Shape of the deformed star}\label{app:A}

The shape (or the figure) of a large celestial body is usually well described by a reference ellipsoid (quadrupolar approximation), i.e. a mathematically-defined  closed quadric surface that is a 3D analogue of an ellipse.
The standard equation of a triaxial ellipsoid centered on the origin of a Cartesian coordinate system $(\X,\Y,\Z)$ and aligned with the axes is
\begin{equation}
\frac{\X^2}{a^2} + \frac{\Y^2}{b^2} + \frac{\Z^2}{c^2} = 1 \ , \llabel{140627b}
\end{equation}
where $a$, $b$, and $c$ are called the semi-principal axes. 
More generally, for any reference frame, equation (\ref{140627b}) can be rewritten as
\begin{equation}
\vv{\R}^T \TA \, \vv{\R} = 1 \ ,
\quad \mathrm{with} \quad \TA = \SR \left[\begin{array}{ccc}  a^{-2} &  0 & 0 \\ 0 &  b^{-2} &0 \\ 0 &  0 & c^{-2} \end{array}\right] \SR^T
\ , \llabel{140716z}
\end{equation}
where 
 $\vv{\R} = (\X,\Y,\Z)$ is a generic point  at the surface of the ellipsoid, and $\SR$ is the rotation matrix that connects the general frame with the aligned axes frame (body frame).
 
Assuming that the deformation of the ellipsoid is small with respect to a sphere of radius $\R_0$, we can express $a=\R_0 (1+\delta_a)$, $b=\R_0 (1+\delta_b)$ and $c=\R_0 (1+\delta_c)$, with $\delta \ll 1$.
Then, we get
\begin{equation}
 \quad \TA \approx \frac{1}{\R_0^2} \, \mathbb{I} - \frac{2}{\R_0^2} \, \SR \left[\begin{array}{ccc}  \delta_a &  0 & 0 \\ 0 & \delta_b &0 \\ 0 &  0 & \delta_c \end{array}\right] \SR^T
\ , \llabel{140716z}
\end{equation}
and the general equation for the ellipsoid becomes
\begin{equation}
(1+\delta_x) \X^2 + (1+\delta_y) \Y^2 + (1+\delta_z) \Z^2 + \delta_1 \X \Y  + \delta_2 \X \Z + \delta_3 \Y \Z = \R_0^2
\ . \llabel{190125a}
\end{equation}

Using previous expression, the radial distance of the ellipsoid surface, $\R = |\vv{\R}| $, can be expressed as
\begin{eqnarray}
\R^2  \!\!\!\!\! & = & \!\!\!\!\!  \vv{\R} \cdot \vv{\R} = \X^2 + \Y^2 +\Z^2 \nonumber \\ 
\!\!\!\!\! & = & \!\!\!\!\!  \R_0^2 - \left( \delta_x \X^2 + \delta_y \Y^2 + \delta_z \Z^2 + \delta_1 \X \Y  + \delta_2 \X \Z + \delta_3 \Y \Z \right)
\ , \llabel{190127a}
\end{eqnarray}
and since all $\delta \ll 1$, we have $\R \approx \R_0 + \DR$, with
\begin{equation}
\DR \approx - \frac{2}{\R_0} \left( \delta_x \X^2 + \delta_y \Y^2 + \delta_z \Z^2 + \delta_1 \X \Y  + \delta_2 \X \Z + \delta_3 \Y \Z \right) \ . \llabel{190127b}
\end{equation}

The deformation of the body is a response to the perturbing potential $V_p$ (Eq.\,(\ref{121026f})).
A very convenient way to define this deformation is through the Love number approach \citep[e.g.][]{Love_1911} 
in which the equilibrium radial displacement $\DR_\eq$ is proportional to the equipotential perturbing surface
\begin{equation}
\DR_\eq = - \hf V_p (\vR) / g \ , \llabel{131021a}
\end{equation}
where $g = G m / \R_0^2$ is the surface gravity, $G$ is the gravitational constant, $m$ is the mass of the star, and $\hf = 1 + \kf$ is the fluid 
second Love number for radial displacement.
In general, for stars we have $\kf \ll 1$, thus $\hf \approx 1$.
Using expression (\ref{130528z}) we get
\be
\DR_\eq \approx - \frac{3}{2 \kf} \left[\hat \vR \cdot \frac{\TI_\eq}{m \R_0} \cdot \hat \vR  \right]  
\ . \llabel{190128a}
\ee
The equilibrium displacements of the star are computed directly from expressions (\ref{151106a})$-$(\ref{151106f}).
However, in a non-equilibrium more general situation, the instantaneous deformation of the star will obey to a similar rheologic law as the one provided by Eq.\,(\ref{130107p}):
\be
\DR + \taub \frac{d}{d t} \DR = \DR_\eq 
\ . \llabel{190129a}
\ee
Thus, the instantaneous deformation of the star is simply given by
\be
\frac{\DR}{\R_0} = - \frac{3}{2 \kf} \left[\hat \vR \cdot \frac{\delta \TI}{m \R_0^2} \cdot \hat \vR  \right]  
\ , \llabel{190128b}
\ee
which allows us to straightforwardly compute the instantaneous shape of the ellipsoid (Eq.\,(\ref{190125a})) from the inertia tensor:
\begin{equation}
\begin{array}{cc}  
\delta_x = \Frac{2 I_{11} - I_{22} - I_{33}}{4 \kf m \R_0^2} \ , &  \delta_1 = \Frac{3 I_{12}}{4 \kf m \R_0^2} \ , \\ 
\delta_y = \Frac{2 I_{22} - I_{11} - I_{33}}{4 \kf m \R_0^2} \ , &  \delta_2 = \Frac{3 I_{13}}{4 \kf m \R_0^2} \ , \\ 
\delta_z = \Frac{2 I_{33} - I_{11} - I_{22}}{4 \kf m \R_0^2} \ , &  \delta_3 = \Frac{3 I_{23}}{4 \kf m \R_0^2} \ .
\end{array}
 \llabel{190128c}
\end{equation}

\section{Initial conditions of the experiments}

For the demonstrations in Sec.~\ref{sec:results} based on observed systems, we provide tables with the initial conditions. These initial conditions and code parameters are also provided in the examples folder in the source code.

\begin{table}
\centering
\begin{tabular}{| c c c c |} 
 \hline
Quantity  & Unit & Earth & Moon \\
 \hline\hline
$\m$             & $\rm{[10^{24}\,kg]}$ & 5.9724 & 0.07346 \\
$R$             & $[\rm{km}]$ & 6371.0 & 1737.4 \\
$\xix$           & & 0.3308 & 0.394 \\
$\kf$           & & 0.933 & 0.0 \\
$\tau$          & $\rm{[s]}$ & [$2^{0}$ - $2^{26}$] & \\
$P_{\rm{spin}}$ & $\rm{[day]}$ & 1 & \\
$a$             & $\rm{[km]}$ & & 0.3844e6 \\
$e$             & & & 0.0 \\
 \hline
\end{tabular}
\caption{Initial conditions for the Earth-Moon system \citep{nasa} in Sec.~\ref{sec:ex1}. We assume that 1) the bodies are initially spherical, 2) Earth's obliquity is zero, 3) the Moon is a rigid body, and 4) the orbit is circular. }
\label{tab:earth_moon}
\end{table}

\begin{table}
\centering
\begin{tabular}{| c c c c |} 
 \hline
Quantity  & Unit & Primary & Secondary \\
 \hline\hline
$\m$             & $\rm{[M_\odot]}$ & 2.39 & 2.33 \\
$R$             & $\rm{[R_\odot]}$ & 2.33 & 2.20 \\
$\xix$           & & 0.052 & 0.052 \\
$\kf$           & & 0.012 & 0.012 \\
$\tau$          & $\rm{[yr]}$ & $10^{-4}$ & $10^{-4}$ \\
$P_{\rm{spin}}$ & $\rm{[day]}$ & 2.55101914403 & 0.728862612579 \\
$a$             & $\rm{[AU]}$ & & 0.3956 \\
$e$             & & & 0.8335 \\
$I$             & $\rm{[deg]} $& & 5.50 \\
$\Omega$             & $\rm{[deg]} $& & 0 \\
$\omega$             & $\rm{[deg]} $& & 36.70 \\
$M$             & $\rm{[deg]} $& & 180 \\
 \hline
\end{tabular}
\caption{Initial conditions for the heartbeat star KOI-54 (Sec.~\ref{sec:ex2}), adopted from \citet{Welsh_2011}. The tidal response parameters are chosen to be in the linear tide regime, while the obliquites are set to zero.  }
\label{tab:koi54}
\end{table}

\begin{table}
\centering
\begin{tabular}{| c c c c |} 
 \hline
Quantity  & Unit & Star & Planet \\
 \hline\hline
$\m$             & $\rm{[10^{30}\,kg]}$ & 2.0088092 & 0.0077459434 \\
$R$             & $[\rm{km}]$ & 702455.0 & 68488.3446 \\
$\xix$           & & 0.070 & 0.25 \\
$\kf$           & & 0 & 0.5 \\
$\tau$          & $\rm{[yr]}$ & 0 & [$10^{-4}$ - $1$]\\
$P_{\rm{spin}}$ & $\rm{[day]}$ & 24.47 & 0.5 \\
$\theta$ & $\rm{[deg]}$ & 0 & [$0$ - $180$] \\ 
$a$             & $\rm{[AU]}$ & & 0.455 \\
$e$             & & & 0.9330 \\
 \hline
\end{tabular}
\caption{Initial conditions for exoplanet HD80606b (Sec.~\ref{sec:ex3}), which are reproduced from \citet{Correia_etal_2014}.  }
\label{tab:hd80606b}
\end{table}

\begin{table}
\centering
\begin{tabular}{| c c c c c |} 
 \hline
Quantity  & Unit & Earth & Moon & Sun \\
 \hline\hline
$\m$             & $\rm{[10^{24}\,kg]}$ & 5.9724 & 0.07346 & 1989000. \\
$R$             & $[\rm{km}]$ & 6371.0 & 1737.4 & 695700 \\
$\xix$          & & 0.3308 & 0.394 & 0.070 \\
$\kf$           & & 0.933 & 0 & 0\\
$\tau$          & $\rm{[s]}$ & 1 & \\
$P_{\rm{spin}}$ & $\rm{[hour]}$ & 5 & \\
$a$             & $\rm{[km]}$ & & 22298.5 & 149598073 \\
$e$             & & & 0.01 & 0 \\
$I$             & $\rm{[deg]}$ & & 0 & 10 \\
 \hline
\end{tabular}
\caption{Initial conditions for the Sun-Earth-Moon experiment in Sec.~\ref{sec:ex4} adopted from \citet{Touma_1998} and \citet{nasa}. Earth's obliquity is zero with respect to the Earth-Moon orbit.  }
\label{tab:sun_earth_moon}
\end{table}

\end{document}